\documentclass[aps,pre,amsmath,amssymb,graphicx,longbibliography,groupedaddress]{revtex4-1} %% DRAFT VERSION - REMOVE THIS IN FINAL VERSION

\usepackage{bm}
\usepackage{ulem}
\usepackage{graphicx}
\usepackage{mdframed}
\usepackage[usenames,dvipsnames]{xcolor}
\usepackage[toc,page]{appendix}
\usepackage{collectbox}
\usepackage{mdframed}
\usepackage{lipsum} % for creating dummy text
\usepackage{lineno} %% DRAFT VERSION - REMOVE THIS IN FINAL VERSION
\newcommand{\mean}[1]{\ensuremath{\left< #1 \right>}}
\newcommand{\de}[1]{\textrm{d}#1}
\newcommand{\pd}[2]{\ensuremath{\frac{\partial #1}{\partial #2}}}
\newcommand{\pdc}[1]{\ensuremath{\frac{\partial}{\partial #1}\,}}
\newcommand{\pddc}[1]{\ensuremath{\frac{\partial^2}{\partial #1 \,^2}\,}}

\begin{document}

\title{Statistical Mechanics of Ecological Systems: Neutral Theory and Beyond}

\author{Sandro Azaele$^{1,*}$, Samir Suweis$^{2,*}$, Jacopo Grilli$^{3}$, Igor Volkov$^{4}$, Jayanth R. Banavar$^4$, Amos Maritan$^{2}$}

\affiliation{$^1$ Department of Applied Mathematics, School of Mathematics, University of Leeds, Leeds LS2 9JT, United Kingdom.
\\ $^2$ Dipartimento di Fisica `G. Galilei' \& CNISM, INFN, Universit\`a di Padova, Via Marzolo 8, 35131 Padova, Italy.\\
$^3$ Department of Ecology and Evolution, University of Chicago, 1101 East 57th Street, Chicago, Illinois 60637, USA.\\
$^4$ Department of Physics, University of Maryland, College Park, Maryland 20742, USA.\\$^*$ These authors contributed equally to this work.}

\begin{abstract}
The simplest theories often have much merit and many limitations, and in this vein, the value of Neutral Theory (NT) has been the subject of much debate over the past 15 years. NT was proposed at the turn of the century by Stephen Hubbell to explain  pervasive patterns observed in the organization of ecosystems. Its originally tepid reception among ecologists contrasted starkly with the excitement it caused among physicists and mathematicians. Indeed, NT spawned several theoretical studies that attempted to explain empirical data and predicted trends of quantities that had not yet been studied. While there are a few reviews of NT oriented towards ecologists, our goal here is to review the quantitative results of NT and its extensions for physicists who are interested in learning what NT is, what its successes are and what important problems remain unresolved. Furthermore, we hope that this review could also be of interest to theoretical ecologists because many potentially interesting results are buried in the vast NT literature. We propose to make these more accessible by extracting them and presenting them in a logical fashion. We conclude the review by discussing how one might introduce realistic  non-neutral elements into the current models. 
\end{abstract}

%\pacs{81.05.Uw,68.37.-d,73.20-r}

\maketitle

\tableofcontents{}

\newpage

%\modulolinenumbers[1] %% DRAFT VERSION - REMOVE THIS IN FINAL VERSION
%\linenumbers %% DRAFT VERSION - REMOVE THIS IN FINAL VERSION

\section{Introduction}
\label{sec:intro}

 \textit{It is interesting to contemplate an entangled bank, clothed with many plants of many kinds, with birds singing on the bushes, with various insects flitting about, and with worms crawling through the damp earth, and to reflect that these elaborately constructed forms, so different from each other in so complex a manner, have been all produced by laws acting around us.}  In this celebrated text from the Origin of Species  \cite{Darwin}, Darwin eloquently conveys his amazement for the underlying simplicity of Nature: despite the striking diversity of shapes and forms, it exhibits deep commonalities that have emerged over wide scales of space, time and organizational complexity. For more than fifty years now, ecologists have collected census data for several ecosystems around the world from diverse communities such as tropical forests, coral reefs, plankton, etc. However, despite the contrasting biological and environmental conditions in these ecological communities, some macro-ecological patterns can be detected that reflect strikingly similar characteristics in very different communities (see BOX 1). This suggests that there are ecological mechanisms that are insensitive to the details of the systems and that can structure general patterns. Although the biological properties of individual species and their interactions retain their importance in many respects, it is likely that the processes that generate such macro-ecological patterns are common to a variety of ecosystems and they can therefore be considered to be universal. The question then is to understand how these patterns arise from just a few simple key features shared by all ecosystems. Contrary to inanimate matter, living organisms adapt and evolve through the key elements of inheritance, mutation and selection. 

This fascinating intellectual challenge fits perfectly into the way physicists approach scientific problems and their style of inquiry. Statistical physics and thermodynamics have taught us an important lesson, that not all microscopic ingredients are equally important if a macroscopic description is all one desires. Consider for example a simple system like a gas. In the case of an ideal gas, the assumptions are that the molecules behave as point-like particles that do not interact and that only exchange energy with the walls of the container in which they are kept at a given temperature. Despite its vast simplifications, the theory yields amazingly accurate predictions of a multitude of phenomena, at least in a low-density regime and/or at not too low temperatures. Just as statistical mechanics provides a framework to relate the microscopic properties of individual atoms and molecules to the macroscopic or bulk properties of materials, ecology needs a theory to relate key biological properties at the individual scale, with macro-ecological properties at the community scale. Nevertheless, this step is more than a mere generalization of the standard statistical mechanics approach. Indeed, in contrast to inanimate matter, for which particles have a given identity with known interactions that are always at play, in ecosystems we deal with entities that evolve, mutate and change, and that can turn on or off as well as tune their interactions with partners. Thus the problem at the core of the statistical physics of ecological systems is to identify the key elements one needs to incorporate in models in order to reproduce the known emergent patterns and eventually discover new ones.

Historically, the first models defining the dynamics of interacting ecological species were those of Lotka \& Volterra, which describe asymmetrical interactions between predator-prey or resource-consumers systems. These models are based on Gause's competitive exclusion principle \cite{gause34}, which states that two species cannot occupy the same niche in the same environment for a long time (see BOX 2). There have been several variants and generalizations of these models, e.g.~\cite{levin1968nonlinear}, yet all of them have several drawbacks: 1) They are mostly deterministic models and often do not take into account stochastic effects in the demographic dynamics \cite{may2001stability}; 2) As the number of species in the system increases, they become analytically intractable and computationally expensive; 3) They have a lot of parameters that are difficult to estimate from ecological data or experiments; 4) It is very difficult to draw generalizations that include spatial degrees of freedom; and 5) While time series of abundance are easily analyzed, it remains challenging to study analytically the macroecological patterns they generate and thus, their universal properties. 

A pioneering attempt to explain macro-ecological patterns as a dynamic equilibrium of basic and universal ecological processes - and that also implicitly introduced the concept of neutrality in ecology - was made by MacArthur and Wilson in the famous monograph of 1967 titled ``The theory of island biogeography'' \cite{MacArthur1967}. In this work, the authors proposed that the number of species present on an island (and forming a local community) changes as the result of two opposing forces: on the one hand, species not yet present on the island can reach the island from the mainland (where there is a meta-community); and on the other hand, the  species already present on the island may become extinct. MacArthur and Wilson's model implies a radical departure from the then main current of thought among contemporary ecologists for at least three reasons: 1) Their theory stresses that demographic and environmental stochasticity can play a role in structuring the community as part of the classical principle of competitive exclusion; 2) The number of coexisting species is the result of a dynamic balance between the rates of immigration and extinction; 3) No matter which species contribute to this dynamic balance between immigration and extinction on the island, all the species are treated as identical. Therefore, at the level of species, they introduced a concept that is now known as neutrality (see BOX 2).

Just a few years later, the American ecologist, H. Caswell, proposed a model in which the species in a community are essentially a collection of non-interacting entities and their abundance is driven solely by random migration/immigration. In contrast to the mainstream vision of niche community assembly, where species persist in the community because they adapt to the habitat, Caswell stressed the importance of random dispersal in shaping ecological communities. Although the model was unable to correctly describe the empirical trends observed in a real ecosystem, it is important because it pictured ecosystems as an open system, within which various species have come together by chance, past history and random diffusion.

Greatly inspired by the theory of island biogeography and the dispersal limitation concept (see BOX 2), in 2001 Hubbell published an influential monograph titled ``The Unified Neutral Theory of Biodiversity and Biogeography''. Unlike the niche theory and the approach adopted by Lotka \& Volterra, the neutral theory (NT) aims to only model species on the same trophic level (monotrophic communities, see BOX 2), species that therefore compete with each other because they all feed on the same pool of limited resources. For instance, competition arises among plant species in a forest because all of them place demands on similar resources like carbon, light or nitrate. Other examples include species of corals, bees, hoverflies, butterflies, birds and so on. The NT is an ecological theory based on random drift, whereby organisms in the community are essentially identical in terms of their per capita probabilities of giving birth, dying, migrating and speciating. Thus, from an ecological point of view, the originality of Hubbell's NT lies in the combination of several factors: $i)$ it assumes competitive equivalence among interacting species; $ii)$ it is an individual-based stochastic theory founded on mechanistic assumptions about the processes controlling the origin and interaction of biological populations at the individual level (i.e. speciation, birth, death and migration); $iii)$ it can be formulated as a dispersal limited sampling theory; $iv)$ it is able to describe several macro-ecological patterns through just a few fundamental ecological processes, such as birth, death and migration \cite{Bell2000,Hubbell2001,Chave2004,butler2009}. Although the theory has been highly criticized by many ecologists as being unrealistic \cite{McGill2003,Nee2005,Ricklefs2006,Clark2009,Ricklefs2012}, it does provide very good results when describing observed ecological patterns and it is simple enough to allow analytical treatment \cite{Volkov2003,Volkov2005,Hubbell2005,Volkov2007,muneepeerakul2008}. However, such precision does not necessarily imply that communities are truly neutral and indeed, non-neutral models can also produce similar patterns \cite{adler2007niche,du2011negative}. Yet the NT does call into question approaches that are either more complex or equally unrealistic \cite{purves2010different,noble2011sampling}. Moreover, NT is not only a useful tool to reveal universal patterns but also it is a framework that provides valuable information when it fails. Accordingly, these features together have made NT an important approach in the study of biodiversity \cite{Harte2003,Chave2004,Hubbell2006,Alonso2006,Walker2007,Black2007,Rosindell2011,Rosindell2012}.

From a physicist's perspective, NT is appealing as it represents a sort of ``thermodynamic'' theory of ecosystems. Similar to the kinetic theory of ideal gases in physics, NT is a basic theory that provides the essential ingredients to further explore theories that involve more complex assumptions. Indeed, NT captures the fundamental approach of physicists, which can be summarized by Einstein's celebrated quote ``Make everything as simple as possible, but not simpler''. Finally, it should be noted that the NT of biodiversity is basically the analogue of the theory of neutral evolution in population genetics \cite{Kimura1985} and indeed, several results obtained in population genetics can be mapped to the corresponding ecological case \cite{Blythe2007}.

Statistical physics is contributing decisively to our understanding of biological and ecological systems by providing powerful theoretical tools and innovative steps \cite{ goldenfeld2007,goldenfeld2010} to understand empirical data about emerging patterns of biodiversity. The aim of this review is not to present a complete and exhaustive summary of all the contributions to this field in recent years - a goal that would be almost impossible in such an active and broad interdisciplinary field - but rather, we would like to introduce this exciting new field to physicists that have no  background in ecology and yet are interested in learning about NT. Thus, we will focus on what has already been done and what issues must be addressed most urgently in this nascent field, that of statistical physics applied to ecological systems. A nice feature of this field is the availability of ecological data that can be used to falsify models and highlight their limitations. At the same time, we will see how the development of a quantitative theoretical framework will enable one to better understand the multiplicity of empirical experiments and ecological data.

This review is organized into five main sections. Sec. \ref{sec:statics} is an attempt to review several important results that have been obtained by solving neutral models at stationarity. In particular, we will present the theoretical framework based on Markovian assumptions to model ecological communities, where different models may be seen as the results of different NT ensembles. We will also show how NT, despite its simplicity, can describe patterns observed in real ecosystems. In Sec. \ref{sec:dynamics}, we will present more recent results on dynamic quantities related to NT. In particular, we will discuss the continuum limit approximation of the discrete Markovian framework, paying special attention to boundary conditions, a subtle aspect of the time dependent solution of the NT. In Sec. \ref{sec:spatial}, we will provide examples of how space plays an essential role in shaping the organization of an ecosystem. We will discuss both phenomenological, and spatially implicit and explicit NT models. A final subsection will be devoted to the modeling of environmental fragmentation and habitat loss. In Sec. \ref{sec:beyond} we will propose some emerging topics in this fledgling field, and present the problems currently being faced. Finally, we will close the review with a section dedicated to conclusions.

%\begin{figure}
%\begin{center}
%\includegraphics[width=19pc]{Fig1-1(hubbel).pdf}
%\caption{[$From$ \cite{Hubbell2001}]. Schematic of the MacArthur and Wilson's theory of island biogeography for
%explaining the number of species on islands as a dynamic equilibrium
%($S^*$) between the rate of immigration of new species onto the island (I = immigration rate)
%and the rate of extinction of species already resident on the island (E = extinction rate).}
%\label{Fig1.1}
%\end{center}
%\end{figure}
%

\begin{figure}
\begin{center}
\includegraphics[width=19pc]{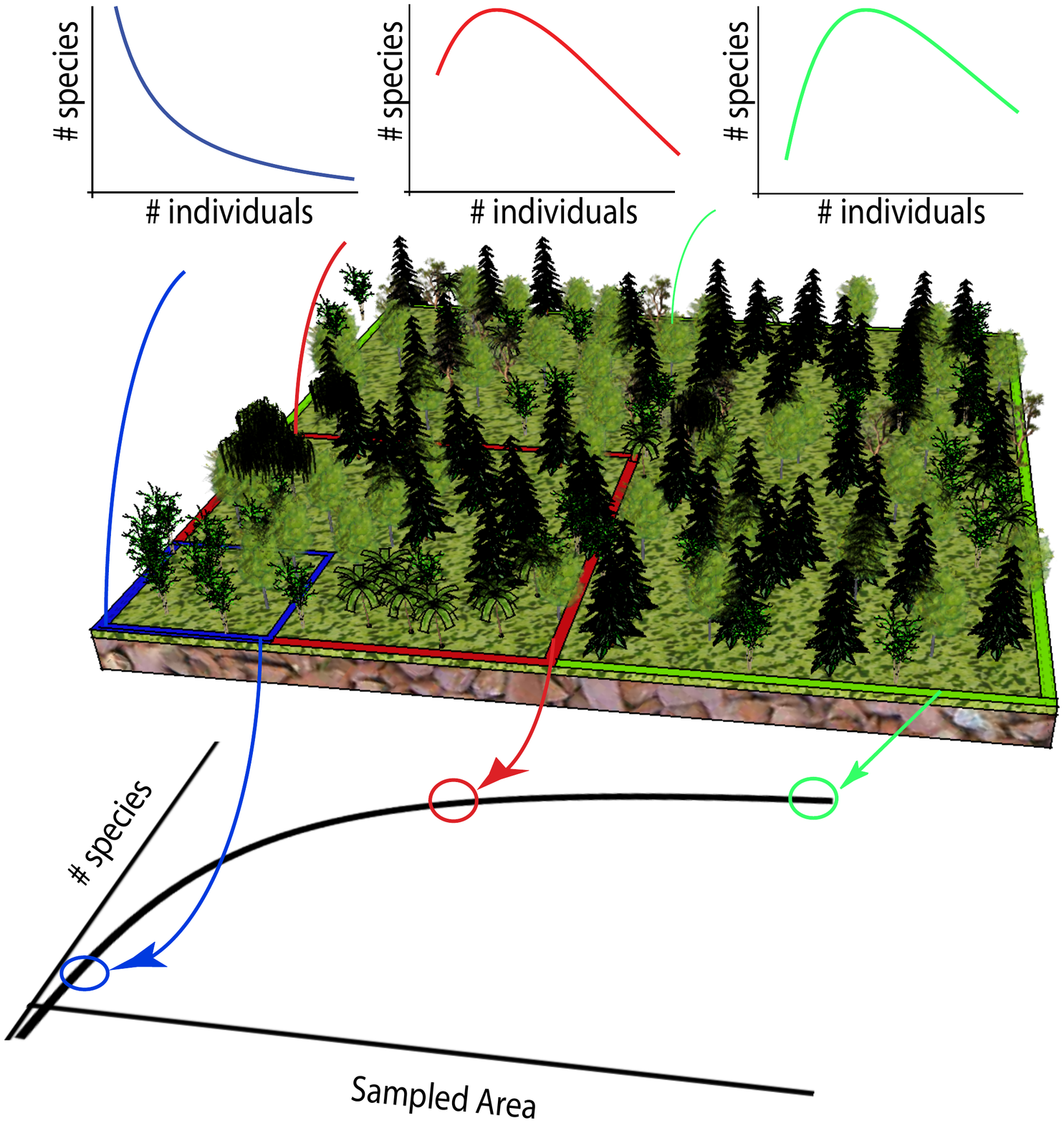}
\caption{[$From$ \cite{Azaele2015}]. Visual Scheme of two important macro ecological patterns (see BOX 1): RSA and SAR. The functional shape of the RSA depends on the spatial scale considered, while the SAR generally displays a tri-phasic behavior (see Section \ref{sec:spatial}). There is a growing appreciation that the various descriptors of biodiversity are intrinsically inter-related, and substantial efforts have been devoted to understand the links between them \cite{Azaele2015}.}
\label{Fig1.2}
\end{center}
\end{figure}

\clearpage
\newpage

\subsection*{BOX 1: Macro-ecological patterns}
\label{sec:box1}
\begin{mdframed}
\begin{itemize}
\item \textbf{Alpha-diversity}: the number of species found in a given area, regardless of their abundance and spatial distributions. Sometimes it is more appropriate to assess the diversity of a local community by taking into account their abundances as well. For this purpose two alternative indexes have been used, the Shannon (H) and Simpson (D) index \cite{simpson1949measurement}
\item \textbf{Beta-diversity}: the probability that two individuals at distance $r$ belong to the same species (i.e., they are conspecific). Several alternative definitions have been used in the literature, including the pair (or two-point) correlation function (PCF), the S\o{}rensen similarity index (SSI) or the Jaccard similarity index (JSI). However, the general purpose of beta-diversity is to describe the turnover of multiple species in space.
\item\textbf{Pair Correlation Function (PCF)}: the correlation in species' abundance between pairs of samples as a function of their distance.
\item \textbf{Relative Species Abundance (RSA)}: the probability that a species has $n$ individuals in a given region. When multiplied by the total number of species in the region, this gives the number of species with $n$ individuals (see section \ref{sec:statics}). This is sometimes called the Species Abundance Distribution (SAD).
\item \textbf{Species (time) Turnover Distribution (STD)}: the probability density function that the ratio of the future to the current population sizes of any species has a value $\lambda$ in a given ecological community (see section \ref{sec:dynamics}).
\item \textbf{Persistence or lifetime distribution (SPT)}: the probability density function of the time interval between the emergence and local extinction of any species within a given area (see section \ref{sec:dynamics}).
\item \textbf{Species Area Relationship (SAR)}: the function that relates the mean number of species, $S$, to the area, $A$, they live in. On a relatively large range of spatial scales it is well approximated by a power law, $S(A)=cA^z$ where $0<z<1$. It has been shown that the SAR in the log-log scale has three qualitatively different behaviours from local to continental spatial scales: approximately linear-like at very small and very large scales, and power-law-like for intermediate scales. This is referred to as a triphasic SAR (see section \ref{sec:spatial}). 
\item \textbf{Endemic Area Relationship (EAR)}: The mean number of species that are present in the area $A$ but not outside it (see section \ref{sec:spatial}).
\end{itemize}
\end{mdframed}

\newpage

\subsection*{BOX 2: Glossary}

\begin{mdframed}
\begin{itemize}
\begin{small}
\item \textbf{Trophic level}: the set of all species belonging to the same level in the food chain. Individuals of species belonging to higher levels feed upon those in the lower ones, while individuals belonging to the same trophic level compete for the same pool of resources. Neutral theory is an ecological theory for species in one specific trophic level, whereas Niche theory (see below) can also deal with species at different trophic levels.
\item \textbf{Gause's competitive exclusion principle}: a pair of species cannot stably co-exist if they feed upon exactly the same resources under the same environmental conditions \cite{gause34}.
\item \textbf{Niche theory}: species can stably coexist in an ecological community if their characteristics (or traits) allow them to specialize on one particular set of resources or environment conditions (niches) in which they are superior to their competitors. In other words, they occupy different niches \cite{hutchinson1957concluding}. Such niche separation is deemed to enhance trade-offs and facilitate co-existence, even though there is no \textit{a priori} method to identify the correct niches that favor co-existence. The underlying rationale of the theory is Gause's exclusion principle, and the classical model is Lotka-Volterra's set of differential equations. In contrast to Niche Theory, Neutral theory claims that niche differences are not essential to co-existence.
\item \textbf{Species-level neutral models}: these models assume that all species are equivalent as they all have the same probability of immigration, extinction and speciation. The only state variable of these models is the number of species in a community and thus, it cannot predict the distribution of the population sizes across species (see RSA below) \cite{MacArthur1967}. %This version of neutrality was present in and Wilson's ``The theory of island biogeography''.
\item \textbf{Individual-level neutral models}: these models assume that all species are equivalent at the individual level,
having the same birth, death, immigration and speciation rates regardless of their identity. Therefore, species
are competitively equivalent. The state variable of these models is the population of any species in a region and
therefore, such models can be used to understand species richness and abundance\cite{Hubbell2001}. 
%This is the new version of neutrality and it was present in Hubbell's ``The Unified Neutral Theory of Biodiversity and Biogeography''.
\item \textbf{Symmetric models}: any model whose outcomes are invariant when exchanging species identities. The family of symmetric models is larger than the neutral ones: for instance, effects of density dependence of individuals on the per capita birth and death rates can be accommodated in symmetric but not neutral models. Some authors do not make a distinction between symmetric and neutral models.
\item \textbf{Dispersal limited process}: any process that constrains offspring to disperse in the vicinity of its parents.
\item \textbf{Multispecies Voter Model with Speciation (MVM)}: a spatially explicit neutral model in which each individual is located in a regular lattice and belongs to a species. Given a spatial configuration at time $ t $, the configuration at time $ t+1 $ is obtained as follows: an individual is chosen at random and is replaced by a copy of one of its nearest neighbours (chosen at random) with a probability $ 1-\nu $, or by an individual belonging to a new species (not already present in the lattice) with probability $ \nu $. $ \nu $ is the speciation parameter, although it may incorporate immigration effects.
\end{small}
\end{itemize}
\end{mdframed}

\newpage

\section{Neutral theory at stationarity}
\label{sec:statics}

Neutral theory deals with ecological communities within a single trophic level, i.e. communities whose species compete for the same pool of resources (see BOX 2). This means that neutral models will generally be tested on data describing species that occupy the same position in the food chain, like trees in a forest, breeding birds in a given region, butterflies in a landscape, plankton, etc.. Therefore, ecological food webs with predator-prey type interactions are not suitable to be studied with standard neutral models. 

\begin{figure}[htbp]
\begin{center}
\includegraphics[width=19pc]{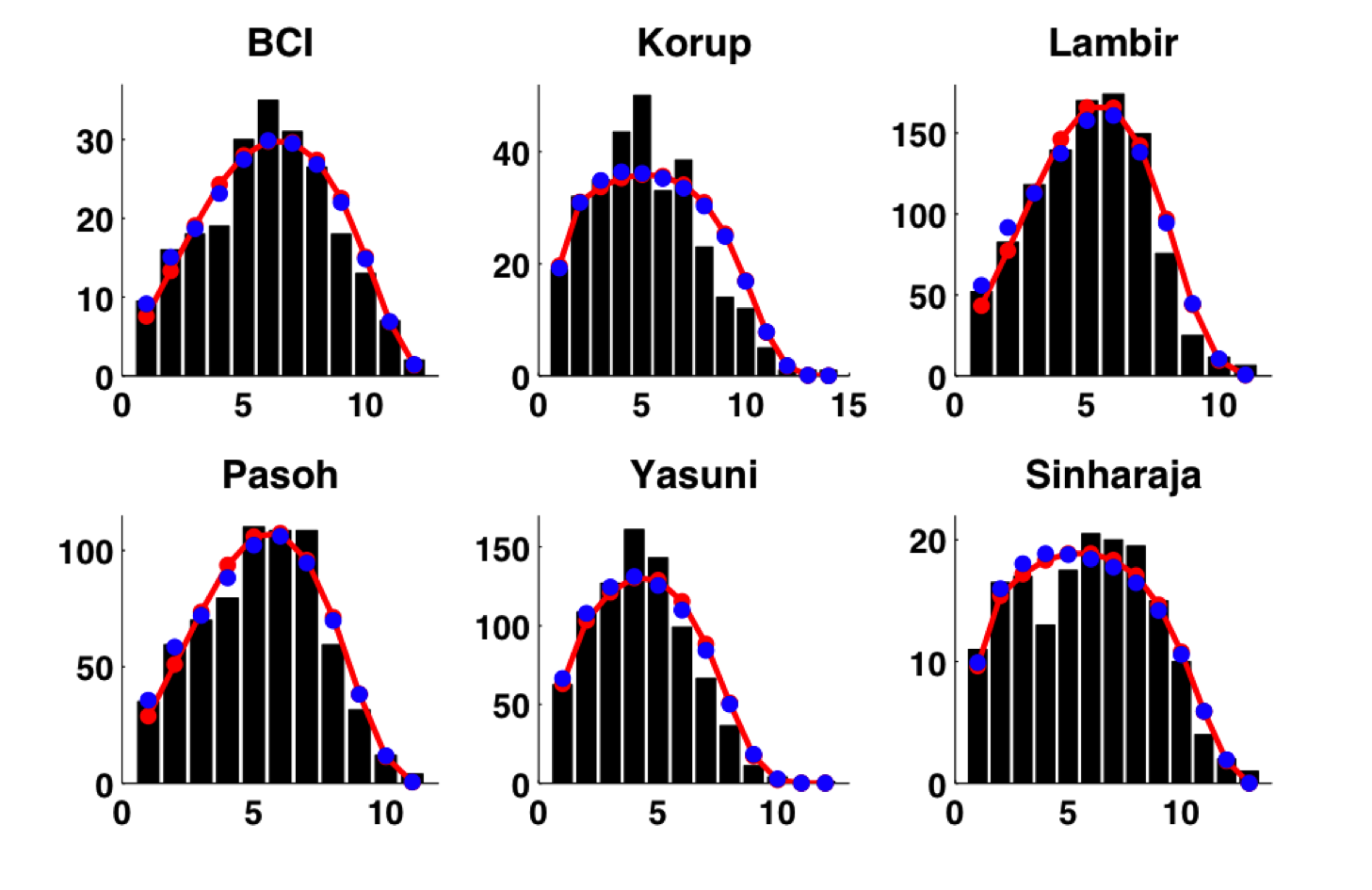}
\caption{[$From$ \cite{Volkov2007}]. Tree relative species abundance
data from the BCI, Yasuni, Pasoh, Lambir, Korup, and Sinharaja plots, for
trees that are $10$ cm in stem diameter at breast height. The
frequency distributions are plotted using Preston's binning method as
described in \cite{Volkov2003} and the bars are the observed number of species binned into $\log (2)$
abundance categories. The
first histogram bar represents $\frac{\langle\phi_1\rangle}{2}$, the
second bar $\frac{\langle\phi_1\rangle}{2}+\frac{\langle\phi_2\rangle}{2}$,
the third bar $\frac{\langle\phi_2\rangle}{
2}+\langle\phi_3\rangle+\frac{\langle\phi_4\rangle}{2}$, the fourth
bar $\frac{\langle\phi_4\rangle}{
2}+\langle\phi_5\rangle+\langle\phi_6\rangle+\langle\phi_7\rangle+\frac{\langle\phi_8\rangle}{
2}$ and so on. Here $\langle\phi_n\rangle$ is the number of species with an abundance $n$.
As examples, we show the fits of a density-dependent symmetric model (red line) and dispersal limitation model (blue circles), which will be studied in greater detail in sec. \ref{sec:statics} and \ref{sec:spatial}, respectively.}
\label{Fig2.1}
\end{center}
\end{figure}

As explained in the introduction, ecologists have been studying an array of biodiversity descriptors over the last sixty years (see BOX 1), including relative species abundance distributions (RSA), species-area relationships (SAR) and  spatial pair correlation function (PCF). For instance, the RSA represents one of the most commonly used static measures to summarize information on ecosystem diversity. The analysis of this pattern reveals that the RSA distributions in tropical forests share similar shapes, regardless of the type of ecosystem, geographical location or the details of species interactions (see Fig. \ref{Fig2.1}). Therefore, the functional form of the RSA (see seminal papers by Fisher \cite{Fisher1943} and Preston \cite{Preston1948} for a theoretical explanation of its origins) has been one of the great problems studied by ecologists. Indeed, a great deal of attention has been devoted to the precise functional forms of these patterns.

It is instructive to start deriving some of these within a very simple but extreme neutral model, which assumes that species are independent and randomly distributed in space. This null model tells us what we should expect when the observed macroecological patterns are only driven by randomness, with no underlying ecological mechanisms. If the density of individuals in a very large region is $ \rho $, then the probability that a species has $ n $ individuals within an area $ a $ is well approximated by a Poisson distribution with a mean $ \rho a $. As defined in Box 1, this is the RSA for the area $ a $.  However, the empirical data are not well described by a Poisson distribution and as we shall see later on, better fits are usually given by log-series, gamma or log-normal distributions. 

The SAR curve can also be calculated within a slightly more accurate model, which still assumes that species are independent and randomly situated in space. Let us now suppose that a region with area $ A_0 $ contains $ S_{tot} $ species in total ($\alpha$-diversity - see Box 1), and that the species $ i $ has $ n_i $ individuals in $ A_0 $. If we consider a smaller area $ A $ within the region, then the probability that an individual will not be found in such area is $ 1-A/A_0 $, while the probability that the whole species $ i $ is not present therein is $(1-A/A_0)^{n_i}=1-p_i$. If we now consider the random variable $ I_i(A) $, which is 1 when the species $ i $ is found within the area $ A $ and 0 if not, then $ \mean{I_i(A)}=p_i=1-(1-A/A_0)^{n_i}$, because $ I_i(A) $ is a Bernoulli random variable with an expectation value $ p_i $. Therefore, the mean number of species in the area $ A $ (i.e. the SAR) is simply $ S(A)=\sum_i \mean{I_i(A)} $, which is 

\begin{equation}
\label{SAR_Coleman} S(A)=S_{tot}-\sum_{i=1}^{S_{tot}} (1-A/A_0)^{n_i}.
\end{equation}
Although this model was originally studied by Coleman \cite{Coleman1981}, we now know that it significantly overestimates species diversity at almost all spatial scales \cite{Plotkin2000}. 

Beta-diversity (see Box 1) can be estimated under the assumptions we have mentioned. Now, regardless of the spatial distance between two individuals, the probability that two of them belong to the same species $ i $ is $ n_i(n_i-1)/[N(N-1)] $, where $ N $ is the total number of individuals in the community, i.e. $ N=\sum_i{n_i} $. Therefore, the probability to find any pair of con-specific individuals is $(1/S^{tot}) \sum_{i=1}^{S^{tot}} n_i(n_i-1)/[N(N-1)] $. This means that the random placement model with independent species predicts that beta-diversity should not depend on the distance between two individuals. Again, we now have clear evidence that the probability that two individuals at distance $r$ belong to the same species is a decaying function of $ r $ \cite{morlon2008general}. 

The failure of the random placement model to capture the RSA, SAR and beta-diversity is a clear indication that ecological patterns are driven by non-trivial mechanisms that need to be appropriately identified. Thus, we shall assess to what extent the NT at stationarity can provide predictions in agreement with empirical data.

There are two related, but distinct analytical frameworks that have been used to mathematically formulate the NT of biodiversity at stationarity for both local and meta-communities. From an ecological point of view, a local community is defined as a group of potentially interacting species sharing the same environment and resources. Mathematically, when modeling a local community the total community population abundance remains fixed. Alternatively, a meta-community can be considered a set of interacting communities that are linked by dispersal and migration phenomena. In this case, it is the average total abundance of the whole meta-community that is held constant. From the physical point of view, these roughly correspond to the micro/canonical (fixed total abundance) and the grand canonical (fixed average total abundance) ensembles, respectively. The micro/canonical ensemble or so-called zero sum dynamics when death and birth events always occur as a pair originates from the sampling frameworks in population genetics pioneered by Warren Evens and Ronald Fisher \cite{Fisher1943}. It should be noted that even though a fixed size sample is one way to analyze available data, for the majority of cases (apart from very small size samples), the grand canonical ensemble approach is that used routinely in statistical physics and it provides a very precise yet largely simplified description of the system. The ultimate reason for this lies in the surprising accuracy of the asymptotic expansion of the gamma function (the mathematical framework heavily uses combinatorials and factorials etc.). The Stirling approximation can be used for very large values of the gamma function, nevertheless, it is quite accurate even for values of the arguments of the order of 20. The advantages of the master equation (see below) and the grand canonical ensemble approach stem from their computational simplicity, which make the results more intuitively transparent.

We now introduce the mathematical tools of stochastic processes that will be used extensively in the rest of the article.

\subsection{Markovian modeling of neutral ecological communities}

\subsubsection{The Master Equation of birth and death}
\label{sec:bdme}

Let $c$ be a configuration of an ecosystem that could be as detailed as the characteristics of all individuals in the ecosystem, including their spatial locations or as minimal as the abundance of a specified subset of species. Let $P(c,t|c_0,t_0)$ be the probability that a configuration $c$ is seen at time $t$, given that the configuration at time $t_0$ was $c_0$ (referred to as $P(c,t)$ for simplicity). For our applications, the configurations $c$ are typically species abundance (denoted by $n$).

Assuming that the stochastic dynamics are Markovian, the time evolution of $P(c,t)$ is given by the Master Equation (ME) \cite{Gardiner1985,VanKampen1992} \begin{equation}
\label{eq:me-def}
\frac{\partial P(c,t)}{\partial t} = \sum_{c'}\bigg(T[ c | c' ] P(c',t) - T[ c' | c ] P(c,t) \bigg),
\end{equation}
where $T[ c | c' ]$ is the transition rate from the configuration $c'$ to configuration $c$. Under suitable and very plausible conditions \cite{VanKampen1992}, $P(c,t)$ approaches a stationary solution at long times, $P_s(c)$, which satisfies the following equation
\begin{equation}
\label{eq:me-st}
\sum_{ c'}\bigg(T[ c | c' ] P_s(c') - T[ c' | c ] P_s(c) \bigg)  = 0.
\end{equation}
This equation is typically intractable with analytical tools, because it involves a sum of all the configurations. If each term in the summation is zero, i.e.
\begin{equation}
\label{eq:me-st-db}
T[ c | c' ] P_s(c') - T[ c' | c ] P_s(c)  = 0 \ ,
\end{equation}
detailed balance is said to hold. A necessary and sufficient condition for the validity of detailed balance is that for \textit{all} possible cycles in the configuration space, the probability of walking through it in one direction is equal to the probability of walking through it in the opposite direction. Given a cycle $\{ c_1, c_2, \dots, c_{n-1}, c_1 \}$, detailed balance holds if and only if, for every such cycle,
\begin{equation}
%\label{eq:dbcondition}
%T[c_2 | c_1 ] T[c_3 | c_2 ] \dots T[c_{n-1} | c_{n-2} ] T[c_1 | c_{n-1} ] =
%T[c_{n-1} | c_1 ] T[c_{n-2} | c_{n-1} ] \dots T[c_{2} | c_{3} ] T[c_1 | c_{2} ]
% \ .
%\end{equation}
\label{eq:dbcondition}
T[c_1 | c_2 ] T[c_2 | c_3 ] \dots T[c_{n-2} | c_{n-1} ] T[c_{n-1} | c_1 ] =
T[c_1 | c_{n-1} ] T[c_{n-1} | c_{n-2} ] \dots T[c_{3} | c_{2} ] T[c_2 | c_{1} ]
 \ .
\end{equation}
This condition evidently corresponds to a time-reversible condition.

Now let us apply these mathematical tools to the study of community dynamics that is driven by random demographical drift. Consider a well-mixed local community. This is equivalent to saying that the distribution of species in space is not relevant, which should hold for an ecosystem with a linear size smaller than, or of the same order as the seed dispersal range. In this case one can use $c= (n_1,\dots, n_S)=\mathbf{n}$ where $n_i$ is the population of the $i$-th species and the system contains $S$ species.
%Since each $n_i=0,\ 1,\ 2\ ,\dots$,$\Omega=\mathbb{N}_0^S$ where $\mathbb{N}_0=\{ 0, 1,\ 2, \dots \}$.
We can rewrite eq.~\ref{eq:me-def} in the following way:
\begin{equation}
\label{eq:me-pop}
\pd{P(\mathbf{n},t)}{t} =  \sum_{ \mathbf{n}'\neq \mathbf{n}}
T[ \mathbf{n} | \mathbf{n}' ] P(\mathbf{n}',t) - \sum_{ \mathbf{n}'\neq \mathbf{n}} T[ \mathbf{n}' | \mathbf{n} ] P(\mathbf{n},t).
\end{equation}
%where $D=\{1,\ 2,\ \dots,\ S \}$.
In this case, $T[ \mathbf{n}' | \mathbf{n} ]$ can take into account birth and death, as well as immigration from a meta-community. The simplest hypothesis is that $T[ \mathbf{n}' | \mathbf{n} ]$ is the result of $S$ elementary birth and death processes that occur independently for each of the $S$ species i.e.
\begin{equation}
\label{MFME}
T[ \mathbf{n}' | \mathbf{n} ] = \sum_{k=1}^S \prod_{i\neq k}\delta_{n_i',n_k}\left(\delta_{n_k',n_k + 1}T_k(n+1|n)+ \delta_{n_k',n_k - 1}T_k(n-1|n)\right)  ,
\end{equation}
where
\begin{equation}
\label{b}  T_{k}(n+1|n) = b(n,k)
\end{equation}
is the birth rate and 
\begin{equation}
\label{d}  T_{k}(n-1|n) = d(n,k),
\end{equation}
the death rate.
This particular choice corresponds
 to a sort of mean-field (ME) approach ~\cite{Volkov2003,Vallade2003,Pigolotti2004,McKane2004a,Alonso2004,Volkov2007,Zillio2008}. 

Our many-body ecological system can also be formulated in a language more familiar to statistical physicists, where we consider the distribution of balls into boxes. The ``boxes" are the species and the ``balls" are the individuals. Birth/death processes correspond to adding/removing a ball to/from one of the boxes using a rule as dictated by Eqs. (\ref{b}) and (\ref{d}).
Eq. (\ref{eq:me-pop}) with the choice (\ref{MFME}) can be simplified if one assumes that the initial condition is factorized as $P(\mathbf{n},t=0)=\prod_{k=1}^S\ P_k(n_k,t=0)$. In this case the solution is again factorized as $P(\mathbf{n},t)=\prod_{k=1}^S\ P_k(n_k,t)$ where each $P_k$ satisfies the following ME in one degree of freedom $n_k$
\begin{equation}\label{1ME}
    \pd{P_k(n_k,t)}{t}= P_k(n_k-1,t)b(n_k-1,k) + P_k(n_k+1,t)d(n_k+1,k) -\left(b(n_k,k)+ d(n_k,k)\right)P_k(n_k,t)\ .
\end{equation}

The stationary solution of eq.(\ref{1ME}) is easily seen to satisfy detailed balance \cite{Gardiner1985,VanKampen1992}.

First, we note that because of the neutrality hypothesis, species are assumed to be demographically identical and therefore, we can drop the $k$ dependence factor from the equations (\ref{b})-(\ref{d}). In other words, we can concentrate on the probability $P(n)$ that a given species (box) has an abundance $n$ (balls). In this case, species do not interact and thus, in our calculation we can follow a particular species (the boxes are taken to be independent). Following equation (\ref{eq:me-st-db}), it is easy to see that the solution of the birth-death ME (\ref{eq:me-def}) that satisfies the detailed balance condition and that thus corresponds to equilibrium, is \cite{Gardiner1985,VanKampen1992}

\begin{equation}\label{P_S}
    P_{s}(n)=P_0\prod_{z=1}^{n}\frac{b(z-1)}{d(z)}
\end{equation}
where $P_0$ can be calculated by the normalization condition $\sum_{n}P_{s}(n)=1$, and assuming that all the rates are positive. In particular, $ b(0)>0 $ and $d(0)=0$.

When there are $S$ boxes, all satisfying the same birth-death rules, the general and unique equilibrium solution is
\begin{equation}\label{P_Smulti}
    P_{s}(n_1,n_2,...,n_{S})=\prod_{k=1}^{S}P_s(n_k)
\end{equation}

Depending on the functional form of $b(n)$ and $d(n)$, one can readily work out the desired $P_s(\mathbf{n})$. We will start with some cases that are familiar to physicists \cite{Volkov2006}, and then move onto more ecologically meaningful cases \cite{Volkov2003,Volkov2007}.

\subsubsection{Physics Ensembles}

\textbf{The Random walk and Bose-Einstein Distribution.} If one chooses $b(n)=b_0$ and $d(n)=d_1$ for $n\geq1$, and $d(n)=0$ otherwise (or alternatively, $b(n)=(n+1)b_0$ and $d(n)=d_1n$), one obtains a pure exponential distribution $P(n)=r^{n}(1-r)$ where $r=b_0/d_1$. Note that in both these cases, $b(n-1)/d(n)$ is the same. Substituting this in equation (\ref{P_Smulti}) gives us the well known Bose-Einstein distribution for non-degenerate energy levels, i.e. 
\begin{equation}\label{P_BE}
   P(n_1,n_2,...,n_{S})=r^{N}(1-r)^{S}, 
\end{equation}
where $N=\sum_k n_k$. Here, $r$ corresponds to $e^{-\beta(\varepsilon-\mu)}$ in the grand canonical ensemble.
   
\textbf{The Fermi-Dirac Distribution}. If $d(n)=d_1n$ and $b(n)=0$ for any $n$ other than 0, and equal to $b_0$ otherwise, then $P(n)=r^n/(1+r)$ for $n=0$ or $n=1$ and $P(n)=0$ for other values of $n$, and accordingly the Fermi-Dirac distribution is achieved
\begin{eqnarray}\label{P_FD}
   P(n_1,n_2,...,n_{S})&=&r^{\sum_k n_k}(1+r)^{-S} \quad\hbox{for $n_k=0$ or $n_k=1$ $\forall i$} \\
   P(n_1,n_2,...,n_{S})&=&0 \quad\hbox{otherwise}.
\end{eqnarray}
   
\textbf{Boltzmann counting} If $d(n)=d_1n$ and $b_n=b_0$ for all $n$, then one obtains a Poisson distribution $P(n)=e^{-r}r^n/n!$, and this leads to 
\begin{equation}\label{P_Boltzman}
   P(n_1,n_2,...,n_{S})\propto \frac{r^{N}}{\prod_{k=1}^{S}n_k!}   .
\end{equation}
This is the familiar grand-canonical ensemble Boltzmann counting in physics, where $r$ plays the role of fugacity. It is noteworthy that, unlike the conventional classical treatment \cite{Huang2001}, where an additional factor of $N!$ is obtained, here one gets the correct Boltzmann counting and thereby avoids the well-known Gibbs paradox \cite{Huang2001}. Thus, if one were to ascribe energy values to each of the boxes and enforce a fixed average total energy, the standard Boltzmann result would be obtained whereby the probability of occupancy of an energy level $\varepsilon$ is proportional to $e^{-\beta \varepsilon}$, where $\beta$ is proportional to the inverse of the temperature.

\subsubsection{Ecological Ensembles}

\textbf{Density independent dynamics.} We now consider the dynamic rules of birth, death and speciation that govern the population of an individual species. The most simple ecologically meaningful case is to consider $d(n)=d n$ and $b(n)=b n$ for $n>0$, and $b(0)>0$, $d(0)=b(-1)=0$. We now define $r=b/d$. Moreover, in order to ensure that the community will not become extinct at longer times, speciation may be introduced by ascribing a non-zero probability of the appearance of an individual from a new species, i.e. $b(0)=b_0=\nu$. In this case the probability for a species of having $n$ individuals at stationarity is:
\begin{equation}\label{P_fisher0}
   P(n)=\tilde\nu[1-\tilde\nu\ln(1-r)]^{-1}r^n/n, 
\end{equation}
where $\tilde\nu=\nu/b$ and $ n>0 $ and $P(0)= 1/(1-\tilde\nu\ln(1-r))$.  
  
The RSA $\langle \phi(n) \rangle$ is the average number of species with a population $n$, and this is simply \cite{Volkov2003}
\begin{equation}\label{P_fisher}
   \langle \phi(n) \rangle =S P(n)= \theta r^n/n.
\end{equation}
This is the celebrated Fisher log-series distribution, i.e. the distribution Fisher proposed as that describing the empirical RSA in real ecosystems \cite{Fisher1943}. The parameter $\theta=S\tilde\nu/[1-\tilde\nu\ln(1-r)]$ is known as the Fisher number or biodiversity parameter.
  
\begin{figure}[htbp]
\begin{center}
\includegraphics[width=19pc]{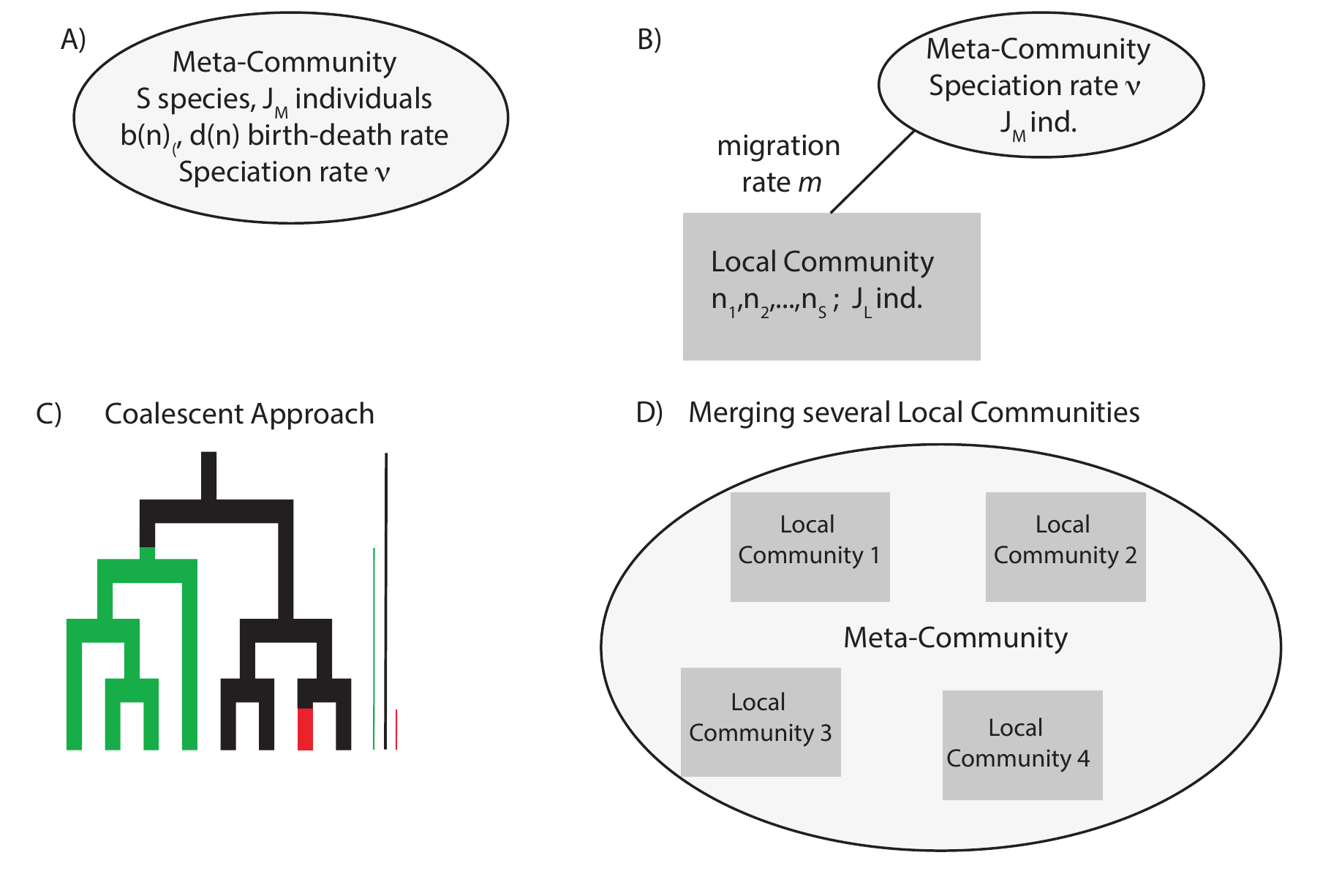}
\caption{Four distinct neutral models of community ecology. In all these models, $n_i$ represents the abundance of the species $i$ in the community, and $J_L$ and $J_M$ the total abundance in the local and meta-communities, respectively. (A) \textit{Hubbell's $zero-sum$ neutral model \cite{Hubbell2001}}. In the local community, each death is immediately followed by a birth or an immigration event. Speciation (or, equivalently, immigration) enables diversity to be maintained in the meta-community. (B) \textit{Local community with immigrants from a meta-community \cite{Vallade2003}}. The local community now interacts with the meta-community through a migration process ($m$). (C) \textit{Coalescent-type approach \cite{Etienne2004}}, where community members are traced back to the ancestors that once immigrated into the community (D) \textit{Joint RSA of many local communities \cite{Volkov2007}}. The whole meta-community RSA distribution is built by considering the joint RSA distributions of multiple local communities}.
\label{Fig2.3}
\end{center}
\end{figure}

In 1948, another great ecologist of the inductive school, F.W. Preston, published a paper \cite{Preston1948} challenging Fisher's point of view. He showed that the Log-series is not a good description for the data from a large sample of birds. In fact, he observed an internal mode in the RSA that was  absent in a Log-series distribution. In particular, Preston introduced a way to plot the experimental RSA data by octave abundance classes (i.e. [$2^{k},2^{k+1}$], for $k=1,2,3,...$), showing that a good fit of the data was represented by a Log-Normal distribution. Indeed, there are several examples of RSA data that display this internal mode feature (see Fig. \ref{Fig2.1}). The intuition of Preston was that the shape of the RSA must depend on the sampling intensity or size of the community. Conventionally, when studying ecological communities, ecologists separate them into two distinct classes: small local communities (e.g., on a island) and meta-communities of much larger communities or those composed of several smaller local communities (see Fig. \ref{Fig2.3}). The neutral modeling schemes for these two cases - that we will denote by the sub-script $L$ and $M$ respectively - are not the same as the ecological processes involved differ. In fact,  the immigration rate ($m$) in a local community is a crucial parameter as the community is mainly structured by dispersal limited mechanisms, and the speciation rate ($\nu$) can be neglected. On the other hand, in a meta-community immigration does not occur (species colonize within the community) and the community is shaped by birth-death-competition processes, although  speciation $\nu$ also plays an important role. A meta-community can also be thought of as consisting of many small semi-isolated local communities, each of which receives immigrants from other local communities. When considering meta-community dynamics, the natural choice is to put a soft constraint on $J_M$, i.e. the total number of individuals is free to fluctuate around the average population $\langle J_M \rangle$. Indeed, one finds that at the largest $J_M$ limit, the results obtained with a hard constraint on $J_M$ are equivalent to those with a soft constraint. On the other hand, when considering a local community, it is safer to place a hard constraint on the total population $J_L$. In both these cases, $S$ represents the total number of species that may potentially be present in the community, while the average number of species observed in the community is denoted by $\langle S \rangle$.
 
There are several ecological meaningful mechanisms that can generate a bell shaped Preston-like RSA. The first of these involves density dependent effects on birth and death rates. The second involves considering a Fisher log-series as the RSA of a meta-community acting as a source of immigrants to a local community embedded within it. The dynamics of the local community are governed by births, deaths and immigration, whereas the meta-community is characterized by births, deaths and speciation. This leads to a local community RSA with an internal mode \cite{McKane2000,Vallade2003,Volkov2003,Volkov2007}. A third way is to incrementally aggregate several local communities (see Appendix A)\\
 
\textbf{Local Dynamics with Density Dependent Birth Rates} One major puzzle in community ecology is the coexistence of a large number of tree species on a local scale in tropical forests. This phenomenon is often explained by invoking density- and frequency dependent mechanisms. Processes that hold the abundance of a common species in check inevitably lead to rare-species advantages, given that the space or resources freed up by density-dependent death can be exploited by less-common species. Therefore, inter-species frequency dependency is the community-level consequence of intra-species density dependence, and thus, they are two different manifestations of the same phenomenon \cite{Volkov2005}.

We begin by noting that the mean number of species with $n$ individuals, $\langle\phi_n\rangle$, is not determined by the absolute rates of birth or death but rather, by their ratio, $r_{i,k}=\frac{b(i,k)}{d(i+1,k)}$. This follows from the observation that $\langle\phi_n\rangle$ is proportional to $\langle r_{1,k}r_{2,k}\ldots r_{n-1,k}\rangle_k$, where the average $\langle\dots\rangle_{k}$ is obtained from all the species. This simple formulation \cite{Volkov2005} is sufficiently general to represent the communities of either symmetric species (in which all the species have the same demographic birth and death rates) or the case of asymmetric or distinct species. The more general asymmetric situation captures niche differences and/or differing immigration fluxes that might arise from the different relative abundances of distinct species in the surrounding meta-community (see BOX 2).

Two of the most prominent hypotheses to explain frequency and density dependence are the Janzen-Connell \cite{janzen1970herbivores,connell1971role} and the Chesson-Warner hypotheses \cite{chesson1981environmental}. These mechanisms generally predict the reproductive advantage of a rare species due to ecological factors and they can be readily captured in a common mathematical framework that will be presented below.

The Janzen-Connell hypothesis postulates that host-specific pathogens or predators act in the vicinity of the maternal parent. Thus, seeds that disperse further away from the mother are more likely to escape mortality. This spatially structured mortality effect suppresses the uncontrolled population growth of locally abundant species relative to uncommon species, thereby producing reproductive advantage to a rare species. The Chesson-Warner storage hypothesis explores the consequences of a variable external environment and it relies on three empirically validated observations: species respond in a species-specific manner to the fluctuating environment; there is a covariance between the environment, and intra- and inter-species competition in function of the abundances of the species; life history stages buffer the growth of population against unfavorable conditions. Such conditions prevail when species have similar per capita rates of mortality but they reproduce asynchronously and there are overlapping generations. 

We now introduce a modified symmetric theory that captures density- and frequency dependence (rare species advantage or common species disadvantage), and in which , $\hat r_n$ should be a decreasing function of abundance in order to incorporate density dependence. The equations of density dependence in the per capita birth and death rates for an arbitrary species of abundance $n$ are: $\frac{b(n)}{n}=b\cdot\left[1+\frac{b_1}{n}+o\left(\frac{1}{n^2}\right)\right]$ and $\frac{d(n)}{n}=d\cdot\left[1+\frac{d_1}{n}+o\left(\frac{1}{n^2}\right)\right]$, for $n>0$ as the leading term of a power series in $(1/n)$, $\frac{b(n)}{n}=b \cdot \sum_{l=0}^\infty b_l n^{-l}$ and $\frac{d(n)}{n}=d\cdot\sum_{l=0}^\infty d_l n^{-l}$, where $b_l$ and $d_l$ are constants. This expansion captures the essence of density-dependence by ensuring that the per-capita rates decrease and approach a constant value for a large $n$, given that the higher order terms are negligible. As noted earlier, the quantity that controls the RSA distribution is the ratio $b_n/d_{n+1}$. Thus, the birth and death rates, $b_n$ and $d_n$, can be defined up to $f(n+1)$ and $f(n)$ respectively, where $f$ is any arbitrary well-behaved function. 

Strikingly, any relative abundance data can be considered as arising from effective density dependent processes in which the birth and death rates are given by the above expressions. Thus, one would expect that the per capita birth rate or fecundity drops as the abundance increases, whereas mortality ought to increase with abundance. Indeed, the per capita death rate can be arranged to be an increasing function of $n$, as observed in nature, by choosing an appropriate function $f$ and appropriately adjusting the birth rate so that the ratio $b_n/d_{n+1}$ remains the same. This then ensures that the RSA does not vary.

The mathematical formulation of density dependence may seem unusual to ecologists familiar with the logistic or Lotka-Volterra systems of equations, wherein density dependence is typically described as a polynomial expansion of powers of $n$ truncated at the quadratic level. Such an expansion is valid when the characteristic scale of $n$ is determined by a fixed carrying capacity. Conversely, here the range of $n$ is from $1$ to an arbitrarily large value and not to some carrying capacity. Therefore, an expansion in terms of powers of $(1/n)$ is more appropriate. For this symmetric model \cite{Volkov2005}, and bearing in mind that $\langle \phi_n\rangle=SP_0\prod_{i=i}^{n-1}\frac{b(i)}{d(i+1)} $, one readily arrives at the following relative species-abundance relationship: 
\begin{equation} 
\label{he1} \langle\phi_n\rangle=\theta\frac{x^n}{n+c}, 
\end{equation} 
where $x=b/d$ and for parsimony, we make the simple assumption that $b_1 =d_1 = c$ and the higher order coefficients, $b_2$, $d_2$, $b_3$, $d_3$, $\dots$, are all 0. The biodiversity parameter, $\theta$, is the normalization constant that ensures the total number of species in the community is $S$, and it is given by $\theta=S\frac{1+c}{c x}/F(1+c,2+c,x)$, where $F(1+c,2+c,x)$ is the standard hypergeometric function. The parameter $c$ measures the strength of the symmetric density- and frequency dependence in the community, and it controls the shape of the RSA distribution. This simple model \cite{Volkov2005} does a good job of matching the patterns of abundance distribution observed in the tropical forest communities throughout the world (see Fig. \ref{Fig2.1}). Note that when $c\rightarrow0$ (the case of no density dependence), one obtains the Fisher log-series. In this case $\theta$ captures the effects of speciation. \\

\textbf{Local community with immigrants from a meta-community.} Various analytical solutions for the Markovian (ME) approach have been suggested in the literature. Based on pioneering work by McKane, Alonso and Sol\'{e} \cite{McKane2000}, an analytical solution was proposed for the ME when the size of both the local community and the meta-community was fixed \cite{McKane2000,Vallade2003,Volkov2003}. Here, we will refer to the work of Vallade and Houchmandzadeh \cite{Vallade2003}, in whose model the birth and death rates in the meta-community are given by
\begin{eqnarray}
\label{bM}  T(n+1|n) &=& b_M(n)= \frac{J_M-n}{J_M}\frac{n}{J_M-1}(1-\nu)  \\
\label{dM}  T(n-1|n) &=& d_M(n)=\frac{n [J_M-n+(n-1)\nu] }{J_M(J_M-1)}.
\end{eqnarray}
Eq. (\ref{bM}) describes the event that a randomly chosen individual of the species under consideration (probability  $n/J_M$) colonizes (probability $1-\nu$) a site occupied by a different species (probability $(J-M-n)/J_M$. Alternatively, Eq. (\ref{dM}) gives the probability that this species is removed (probability $n/J_M$) and replaced by another species from within (with probability $(1-\nu)(J_M-n)/(J_M-1)$) or outside (migration event with probability $\nu$) the system. Eqs. (\ref{bM})-(\ref{dM}) also correspond to the transition rates for the mean field Voter Model (see MVM Box 2 and Section IV).

Let $\langle \phi_n(t) \rangle_M$ designate the average number of species of abundance $n$ in the meta-community. Then, in the stationary limit ($t\rightarrow\infty$), the species that contribute to $\langle \phi_n(t) \rangle_M$ are those which entered the system by mutation at time $t-\tau$ and that have reached size $n$ at time $t$ \cite{Vallade2003,Suweis2012}. As speciation is a Poissonian event (of rate $\nu$) and due to neutrality, all species have the same probability $p(\tau)d\tau=\nu d\tau$ of appearing between the time interval $[\tau, \tau +d\tau]$. Thus, the time evolution of $\langle \phi_n(t) \rangle$ is then simply given by $\langle \phi_n(t) \rangle=\int_0^tP(n,t-\tau)p(\tau)d\tau=\nu\int_0^tP(n,\tau)d\tau$, where $P(n,t)$ is the solution of the ME (\ref{eq:me-pop}) with transition rates given by equations (\ref{bM})-(\ref{dM}) and with the initial condition $P(n,0)=\delta_{n,1}$ ($\delta$ represents the Kronecker delta). Using Laplace transforms, the stationary RSA for the meta-community of size $J_M$ is \cite{Vallade2003}:
\begin{equation}
\langle\phi_k\rangle_M=\frac{\theta\Gamma(J_M+1)\Gamma(J_M+\theta-k)}
{k\Gamma(J_M+1-k)\Gamma(J_M+\theta)},
\end{equation}
where $\theta=(J_M-1)\nu/(1-\nu)$ .

Therefore the density of species of relative abundance $\omega=k/J_M$ in the meta community can be written as:
 \begin{equation}
g_M(\omega)=\frac{\theta(1-\omega)^{\theta-1}}{\omega},
\end{equation}
where $g_M(\omega)=\lim_{J_M\rightarrow\infty}J_M\langle\phi_k\rangle_M$ and with  $\omega$ and $\theta$ kept fixed (implying that $\nu$ tends to zero and $k$ tends to infinity).

Let us now consider a local community (of size $J_L$) in contact with the aforementioned meta-community. As $J_M \gg J_L$, then mutations within the local community can be neglected ($\nu\rightarrow0$) and the equilibrium distribution in the meta-community is negligibly modified by migration from or towards the community. Nevertheless, there is migration from the meta-community to the local community and thus, a migration rate $m$ must be included in the transition rates for the local community dynamics:
\begin{eqnarray}
\label{bL} b_L(n)& =& \frac{J_L-n}{J_L}\bigg[\frac{n}{J_L-1}(1-m)+m\omega\bigg] \\
\label{dL} d_L(n)& =& \frac{n}{J_L}\bigg[\frac{J_L-n}{J_L-1}(1-m)+m(1-\omega)\bigg],
\end{eqnarray}
where $\omega=k/J_M$ is the relative abundance of the species in the meta-community and $k$ is the abundance of the tracked species within the meta-community.

The stationary solution $P_n(\omega)$ of the corresponding ME can be calculated exactly using Eq. (\ref{P_S}), considering $m$ as the immigration rate \cite{Vallade2003}:

\begin{equation}
P_n(\omega)=\binom{J_L}{n}\frac{(\mu\omega)_n[\mu(1-\omega)]_{J_L-n}}{(\mu)_{J_L}},
\end{equation}
where $\binom{J_L}{n}$ is the binomial coefficient, $(a)_n=\Gamma(a+n)/\Gamma(a)$ is the Pochammer coefficient, and $\mu=(J_L-1)m/(1-m)$.

Finally, the average number of species with abundance $n$ in the local community, $\langle\phi_n\rangle_L$, is given by the following expression:
\begin{equation}
\langle\phi_n\rangle_L=\sum_{k=1}^{J_M}
P_n(k/J_M)\langle\phi_k\rangle_M,
\end{equation}
which displays an internal mode for appropriate values of the model parameters. A generalization of this result to a system of two communities of arbitrary yet fixed sizes that are subject to both speciation and migration \cite{Vallade2006} has also been carried out.

Alonso and McKane \cite{Alonso2004} suggested that the log-series solution for the species abundance distribution in the meta-community is applicable only for species-rich communities, and that it does not adequately describe species-poor meta-communities. An application of the ME approach to the asymmetric species case was considered \cite{Alonso2008} and it was further demonstrated that the species abundance distribution has exactly the same sampling formula for both zero-sum and non-zero-sum models within the neutral approximation \cite{Etienne2007b,Haegeman2008}. The simplest mode of speciation -- a point mutation -- has been the one most commonly used to derive the species abundance distribution of the meta-community. A Markovian approach incorporating various modes of speciation such as random fission was also presented by Haegeman and Etienne \cite{Haegeman2009,Haegeman2010}. \\

\subsection{Coalescent-type approach of Neutral Theory}
An exciting approach pioneered by Etienne and Olff, as well as by Alonso, considers a coalescent-type phenomenon where community members are traced back to their ancestors that once immigrated into the community \cite{Etienne2004,Etienne2005b}. A succinct comparison of the two different approaches adopted is presented in \cite{Etienne2005a,Chave2006}.

Within this framework, the local community consists of a fixed number of individuals $J$ that undergo zero-sum dynamics. At each timestep a randomly chosen individual dies and is replaced by a new individual that is an offspring of a randomly chosen individual in the community (of size $J-1$), or by one of $I$ potential immigrants from the meta-community. The immigration rate is thus $m=I/(I+J-1)$, where the parameter $I$ is merely a proxy for the immigration rate. The meta-community is governed by the same dynamics, with speciation replacing immigration.

Two observations are crucial to derive the sampling formula for the species abundance distribution. Firstly, because there is no speciation in the local community, each individual is either an immigrant or a descendent of an immigrant. Thus, the information pertaining to the species abundance distribution is specified by considering ``the ancestral tree" (tracing back each individual to its immigrant ancestor - which is somewhat similar to the phylogenetic tree construction in genetics) and the species composition of the set of all the ancestors. Secondly, this set of all ancestors can be considered as a random sample from the meta-comunity and its species abundance distribution is provided by the well-known multivariate Ewens distribution, corresponding to the Fisher's log-series in the limit of large sample size, that describes neutral population genetics models with speciation and no immigration. As we do not know the ancestor of each individual, the final formula for the species-abundance composition of the data comes from summing the probabilities of all possible combinations of ancestors that give rise to the observed data.

In a tour de force calculation, the resulting sampling formula is found to be \cite{Etienne2005a}
\begin{equation}
	P[D|\theta,m,J]=\frac{J!}{\prod_{i=1}^Sn_i\prod_{j=1}^J\Phi_j!}
	\frac{\theta^S}{(I)_J}\sum_{A=S}^JK(D,A)\frac{I^A}{(\theta)_A},
\end{equation}	
where $P[D|\theta,m,J]$ is the probability of observing the empirical data $D$ consisting of $S$ species with 
abundances $n_i$, $i\in[1,S]$ given the value of the fundamental biodiversity number $\theta$ of the meta-community, the immigration rate $m$ and the local community size $J$. $\Phi_j$ is the number of species with abundance $j$,
\begin{equation}
	K(D,A)=\sum_{\{a1,\dots,a_S|\sum_{i=1}^Sa_i=A}\prod_{i=1}^S
	\frac{\bar s(n_i,a_i)\bar s(a_i,1)}{\bar s(n_i,1)},
\end{equation}
where $\bar s$ is the unsigned Stirling number of the first kind.
Finally, $(x)_y$ is the Pochhammer symbol $\prod_{i=1}^y(x+i-1)$. 

These expressions are simplified versions of the original results presented by the authors and as expected, they provide a comparable fit to the ME approach for the data from the tropical forests. It is heartening that equivalent results can be derived using very different approaches.

\section{Dynamics in neutral theory}
\label{sec:dynamics}

So far we have focused on the implications of neutral theory when models describing neutral patterns reach a steady state. The stationary condition allows one to take advantage of a variety of different mathematical techniques to obtain analytical expressions of ecological patterns. However, stationarity is not always a good approximation, either because the ecosystem is still in a state of flux, or because the assumption may hide different and important processes that lead to the same final steady state. Furthermore, one can calculate time dependent correlation functions \cite{hashitsume1991statistical} (see, for example, (\ref{delta}) below). It is therefore essential to understand the temporal behavior of ecosystems in order to discriminate between ecological processes that would otherwise be indistinguishable at stationarity. Statistical comparisons between time-dependent patterns are usually more difficult than those between stationary patterns because they require more data and long empirical time series that are rarely available. In addition, although time-dependent solutions facilitate stronger tests when confronted with data, they are more difficult to obtain and this is the reason why only a few studies have investigated the temporal behavior of neutral models.

An important method to study the time dependence is van Kampen's system size expansion. Assuming that the total population of the system, $N$, is very large and the initial population is of the order of $N$, one expects that, at early times at least, the probability distribution of the system, $ P(n,t) $, would peak sharply around the macroscopic value $ n=N\phi(t) $ with a width of order $ \sim \sqrt{N} $. The function $ \phi(t) $ is chosen to describe the evolution of the peak. These considerations lead to the Ansatz $ n=N\phi(t)+\sqrt{N}\xi $, where $ \xi $ is a new random variable. When expressing the ME in terms of $ \xi $ and expanding it in $ \sqrt{N} $, at order $ \sqrt{N} $ one recovers the deterministic evolution of the macroscopic state of the system, and at order $ N^0 $ one obtains a linear Fokker-Planck equation whose coefficients depend on time through the function $ \phi(t) $. Therefore, fluctuations around the peak are Gaussian in a first approximation (details of the method can be found in Ref.\cite{VanKampen1992}). This expansion has some attractive features but it is usually only a good approximation of the temporal evolution of the original ME at short or intermediate temporal scales, since the Gaussian behavior is usually lost over longer periods. The time dependence of mean-field neutral models was investigated by means of van Kampen's system size expansion in \cite{McKane2000} and \cite{McKane2004a} and a good agreement with simulations was found at early times. 

Some neutral models can also be formulated in terms of discrete-time Markov processes that have the form $ p_i(t+1)=\sum_{j}Q_{ij}p_j(t) $, where $ p_i(t) $ is the probability that a species has $ i $ individuals at time $ t $, and $ Q_{ij} $ is the probability that a species changes its abundance from $ j $ to $ i $ in one time step (i.e. the transition matrix). If the community has $ N $ individuals in total and either one birth or one death is allowed in one time step, then $ Q $ is a tridiagonal matrix with dimension $ N+1 $. Solving these models in time is basically equivalent to finding the eigenvalues along with the left and right eigenvectors of the matrix $ Q $, which is usually a non-trivial task. Using a theorem derived in population genetics, Chisholm \cite{Chilsom2011} was able to obtain the eigen decomposition of the transition matrix of a neutral model that describes the behavior of the local community in the original Hubbell model \cite{Hubbell2001}. Also, he was able to show that the first two eigenvectors are sufficient to provide a good approximation of the full time-dependent solution.

Other authors have focused on specific temporal patterns that can sometimes be calculated more easily. For instance, a measure that has been used to quantify the diversity of a community is the Simpson index \cite{simpson1949measurement}, $ D $, which is defined as the probability that two individuals drawn randomly from a well-mixed community belong to the same species. If the individuals are drawn with replacement, then one obtains $ D=\sum_i n^2_i/N^2 $, where $ n_i $ is the number of individuals of species $ i $ (we have already encountered this index in Section II when we calculated the beta-diversity in the random placement model). This index is also a good indicator of the relative importance of spatial effects in a community. The dynamic evolution of $ D $ is usually simple enough to allow analytical calculations
\cite{vanpeteghem2008dynamics}. Other interesting dynamic properties of neutral models have been studied, where species' extinction and monodominance were investigated by looking into the first-passage properties and fixation times of the process \cite{babak2006continuous,Babak2009}. Empirical power-law relationships between the temporal mean and variance of population fluctuations (the so-called Taylor's power law) are also in good agreement with neutral predictions \cite{Keil2010}.

Although it is not mandatory from the neutral assumption, models often assume that demographic stochasticity is the main source of fluctuations in stochastic neutral dynamics \cite{ricklefs2003comment,Hu2006}. Demographic randomness originates from the intrinsic stochastic nature of birth and death events within a discrete population of individuals. However, other sources may be important, such as environmental stochasticity that, by contrast, encompasses effects of abiotic and biotic environmental variables on all individuals. There is theoretical and empirical evidence that the dynamics of the more common species may be driven by environmental stochasticity \cite{lande2003stochastic}. Consequently, neutral models only based on demographic stochasticity will overestimate the expected times to extinction for abundant species \cite{Ricklefs2006}, whose temporal fluctuations will also be underestimated \cite{mutshinda2008species}. Incorporating an environmental source of randomness as well as more realistic forms of speciation have made some dynamic aspects of NT more realistic \cite{allen2007setting}.

Here we will focus on the dynamic aspects of the species abundance distribution (SAD) \cite{Azaele2006} which, under appropriate assumptions, can be studied in detail \cite{lehnigk1993generalized,Masoliver2012} and whose predictions can be benchmarked against empirical data. We will also review recent progress in modeling dynamic patterns, including the species turnover distribution \cite{Azaele2006} as well as species' persistence-times (or lifetimes) distributions \cite{pigolotti2005species,Bertuzzo2011,pinto2011,suweis2012species} (see BOX 1).

\subsection{The continuum limit of the Master Equation}

The microscopic description of a system starts by correctly identifying the variables that define all the possible configurations of the system. Having decided how to describe the states of the system, the next step is to consider the transition rates among different states. The configurations of an ecological community can be described by different variables according to different levels of coarsening of the spatio-temporal scales. However, for the sake of simplicity, we will focus on ecosystems comprising $ S $ species and $ N $ individuals in total, in which the configurations are specified fully by the variable $\mathbf{n}=(n_1,\dots, n_S)$, where $n_i$ is the population of the $i$-th species. Therefore, for the time being, we will ignore any spatial effects, which will be considered in section \ref{sec:spatial}.

If the ecological community at time $ t_0 $ was in the configuration $ \mathbf{n}_0 $, then the probability that the system will be in configuration $ \mathbf{n} $ at time $ t>t_0 $ is  $P(\mathbf{n},t|\mathbf{n}_0,t_0) $ ($ P(\mathbf{n},t) $ for brevity). Assuming a Markovian stochastic dynamics, the evolution of $ P(\mathbf{n},t) $ is governed by the ME (\ref{eq:me-pop}) with transition rates $T[ \mathbf{n} | \mathbf{n}' ]$ from the states $ \mathbf{n}' $ to $ \mathbf{n}$. As seen in section II, the specification of a model is tantamount to defining the transition rates. This equation can be recast in another form that allows a simple expansion. Interactions among species change the variable $ \mathbf{n} $ in steps of size $ \mathbf{r}_\ell $, where $ \ell=1,2,...,L $ and $ L $ is the total number of distinct possible changes or reactions. For instance, if a species can either decrease or increase its population by $k$ in a given time step, then $ L=2 $. Species that change by one or that remain constant in a given time interval are instead described by $L=3$, and when the total population is conserved, we have $ \sum\limits_{\ell=1}^{L}r_{i,\ell}=0 $. Of all the possible reactions, however, only a few are usually significant, ecologically relevant or meaningful. Moreover, unlike chemical reactions these meaningful reactions can be considered irreversible. Therefore, instead of summing over states $ \mathbf{n}' $, as in eq.(\ref{eq:me-pop}), we can sum over $L$ different reactions. If $ t_{\ell}[\mathbf{n}] $ is the transition rate of the $ \ell $-th reaction which involves the jumps $  \mathbf{r}_\ell $ when species' populations are $ \mathbf{n} $, then we can recast eq.(\ref{eq:me-pop}) in the form

\begin{equation}
\label{eq:me-pop2}
\pd{P(\mathbf{n},t)}{t} =  \sum_{\ell=1}^{L}\big \{
t_{\ell}[ \mathbf{n} - \mathbf{r}_\ell ] P(\mathbf{n} - \mathbf{r}_\ell,t) - t_{\ell}[ \mathbf{n} ] P(\mathbf{n},t) \big \}
 \ .
\end{equation}

Assuming that $ t_{\ell} [ \mathbf{n} ] $ is only a function of $ \mathbf{x}=\mathbf{n}/N $ (this corresponds to eqs.(\ref{bM}), (\ref{dM}) with $ N=J_M $ and (\ref{bL}), (\ref{dL}) with $ N=J_L $ and $ N\gg 1$ in both cases), we can write 
\begin{equation}
\label{eq:me-pop3}
\pd{P(\mathbf{x},t)}{t} =  \sum_{\ell=1}^{L}\big \{
t_{\ell}\left[ \mathbf{x} - \frac{\mathbf{r}_\ell}{N} \right] P\left(\mathbf{x} - \frac{\mathbf{r}_\ell}{N},t\right) - t_{\ell}\left[ \mathbf{x} \right] P(\mathbf{x},t) \big \}
 \ ,
\end{equation}
where $ \mathbf{n} $ has been replaced by $ N\mathbf{x} $. This form of the ME suggests that $ |\mathbf{r}_\ell| /N \ll 1 $ when $ N\gg 1 $ and hence, a Taylor expansion around $  \mathbf{x} $ of $ t_{\ell} $ and $ P $ is legitimate, at least far from the boundaries. This is known as the Kramers-Moyal expansion \cite{Gardiner1985}. Truncating to  second order in $ N $, we obtain a Fokker-Planck equation of the form
\begin{equation}
\label{eq:multi-FP}
\pd{P(\mathbf{x},\tau)}{\tau}= - \sum_{i=1}^{S}\pdc{x_i}[A_i(\mathbf{x})P(\mathbf{x},\tau)] + \dfrac{1}{2N}\sum_{i,j=1}^{S}\dfrac{\partial^2}{\partial x_i\partial x_j}[B_{ij}(\mathbf{x})P(\mathbf{x},\tau)]
+O(N^{-2})\ ,
\end{equation}
where the time has been rescaled ($ \tau=t/N $), the functions $ A_i(\mathbf{x}) $ have been defined as follows
\begin{equation}
\label{eq:FP-A}
A_i(\mathbf{x})\equiv \sum_{\ell=1}^{L}r_{i,\ell}t_{\ell}\left[ \mathbf{x} \right]
\,
\end{equation}
and the entries of the matrix $ B(\mathbf{x}) $ are
\begin{equation}
\label{eq:FP-B}
B_{ij}(\mathbf{x})\equiv \sum_{\ell=1}^{L}r_{i,\ell}r_{j,\ell}t_{\ell}\left[ \mathbf{x} \right]
\ .
\end{equation}

The matrix $ B(\mathbf{x}) $ is a $ S\times S $ symmetric matrix that is positive semi-definite \cite{Gardiner1985}. For one-step jumps, the entries of the matrices $ \mathbf{r}_{\ell}\otimes\mathbf{r}_{\ell} $ can only be 1, -1 or 0. Unlike the ME (\ref{eq:me-pop2}), the Fokker-Planck equation (\ref{eq:multi-FP}) governs the evolution of \textit{continuous} stochastic variables that represent the relative species' populations under the so-called diffusion approximation. As the total population extends to infinity, we recover the deterministic evolution given by the Liouville equation

\begin{equation}
\label{eq:multi-Liouville}
\pd{P(\mathbf{x},\tau)}{\tau} = - \sum_{i=1}^{S}\pdc{x_i}[A_i(\mathbf{x})P(\mathbf{x},\tau)]
\ ,
\end{equation}
whose solution has the simple form $ P(\mathbf{x},\tau)=\delta(\mathbf{x}-\mathbf{x}(\tau)) $, where $ \delta $ is the Dirac delta function. This form of $ P $ implies that $ \mathbf{x}(\tau) $, i.e. the relative species' populations, are the solutions of the deterministic system $\dot{x}_i(\tau)= A_i(\mathbf{x}) $, where $ i=1,2,\ldots,S $ \cite{Gardiner1985}.

So far we have assumed that the system has a fixed and finite total population, $ N $, and therefore each population cannot exceed $ N $, despite its fluctuations. However, sometimes it is useful to relax this constraint and allow each population to fluctuate within the interval $ [0,\infty) $. In this case, a parameter may or may not exist that allows one to recover the deterministic limit. However, if we assume that $ t_{\ell}\left[ \mathbf{n} \right] $ are smooth functions of $ \mathbf{n} $, and when $ \mathbf{n} \longrightarrow \mathbf{n}+\mathbf{r}_{\ell}$, they vary little with respect to a characteristic scale of the model (e.g., the total average population), and then, we can treat $  \mathbf{n}  $ as continuous variables and expand $ t_{\ell}[ \mathbf{n} - \mathbf{r}_\ell ] P(\mathbf{n} - \mathbf{r}_\ell,t) $ around $ \mathbf{n} $. Starting from the ME (\ref{eq:me-pop2}) and using $ \mathbf{x} $ to indicate the continuous populations, we obtain 

\begin{equation}
\label{eq:multi-FP2}
\pd{P(\mathbf{x},t)}{t} = - \sum_{i=1}^{S}\pdc{x_i}[A_i(\mathbf{x})P(\mathbf{x},t)] + \dfrac{1}{2}\sum_{i,j=1}^{S}\dfrac{\partial^2}{\partial x_i\partial x_j}[B_{ij}(\mathbf{x})P(\mathbf{x},t)]
\ ,
\end{equation}
where time has not been rescaled and $ A_i(\mathbf{x}) $ and $ B_{ij}(\mathbf{x}) $ are as in Eqs. (\ref{eq:FP-A}), (\ref{eq:FP-B}). If we are within a neutral framework with one-step jumps, then  $ S=1, L=2 $ and $ r_1=1,r_2=-1 $. Setting $ t_1[x]\equiv b(x) $ (birth rate) and $  t_2[x]\equiv d(x) $ (death rate), from Eq. (\ref{eq:multi-FP2}) one gets
\begin{equation}
\pd{P(x,t)}{t} = \pdc{x}[d(x)-b(x)]P(x,t) + \frac{1}{2}\pddc{x}[d(x)
+ b(x)]P(x,t)\label{eq:FP}\ .
\end{equation}

Starting from the discrete formulation of the birth and death rates that we have seen previously, one can derive the expressions of the corresponding continuous rates. Following our earlier discussion, we can write the following general expansion 

\begin{subequations}
\begin{align}
\frac{b(x)}{x} = b_1 + \frac{b_0}{x} + \cdots \\
\frac{d(x)}{x} = d_1 + \frac{d_0}{x} + \cdots
\end{align}\label{para}
\end{subequations}
where we have dropped higher order terms in $1/x$. Clearly, Eq. (\ref{para}) can be applied when species have sufficiently large populations. The constants $b_0$ and $d_0$ produce a density dependence effect that causes a rare species advantage (disadvantage) when $b_0>d_0$ ($b_0<d_0$). This effect has its roots in ecological mechanisms, such as the Janzen-Connell \cite{janzen1970herbivores,connell1971role} or Chesson-Warner effects \cite{chesson1981environmental}. However, the presence of net immigration or speciation in a local community \cite{Volkov2003} can also produce such density dependence, which is captured here by a mean field approach. Usually, the skewness of the RSA of various tropical forests and coral reefs indicates a rare species advantage \cite{Volkov2005}, and thus, we will use $b_0 > d_0$. In addition, to simplify the analytical treatment and for parsimony, we choose $b_0=-d_0$. Substituting the rates in (\ref{para}) into Eq. (\ref{eq:FP}) and setting $b_0=-d_0>0$, we obtain
\begin{equation}
\pd{P(x,t)}{t} = \pdc{x}[(\mu x -b)P(x,t)] + D\pddc{x}[xP(x,t)]
\label{fokker}\ ,
\end{equation}
where $ D=(d_1+b_1)/2$, $ \mu=1/\tau =d_1-b_1 > 0 $ and $ b=2b_0 >0 $. $P(x,t)\equiv P(x,t|x_0,0)$ is the probability density function of finding $x$ individuals in the community at time $t$, given that at time $t=0$ there were $x_0$. Therefore $\int_n^{n+\Delta n}P(x,t)\de{x}$ is the fraction of species with a population between $n$ and $n+\Delta n$ at time $t$. Eq. (\ref{fokker}) defines our model for species-rich communities and it governs the time evolution of a community population under the neutral approximation. The deterministic term drives the population to the stationary mean population per species $b/\mu$ and therefore, within our model, the population and the number of species can also fluctuate at stationarity with the ratio of the population to the number of species being fixed on average. Thus, this model is more flexible than the original model by Hubbell
\cite{Hubbell2001} in which `zero-sum dynamics' that fixes the total number of individuals in a community was assumed.

The link established between the FP equation and the ME provides a useful interpretation of the coefficients: $\mu=d_1-b_1$ is the imbalance between the per capita death and birth rates that inexorably drives the ecosystem to extinction in the absence of immigration or speciation. In this model $\tau$ is the characteristic time scale associated with perturbations to the steady-state. When $D\tau \gg 1$ (as for the tropical forests we have analyzed) we have $d_1=(2D+\tau^{-1})/2\simeq D$, and so $D$ can be thought of as an individual death rate. Finally, $b$ plays the role of an effective immigration (or speciation) rate that prevents the community from becoming extinct.

As expected, different choices of $b_0$ and $d_0$ lead to very good fits of the RSA of various tropical forests. In particular, it is possible to see that the RSA fits are readily improved for large $x$ when $b_0$ and $d_0$ are arbitrary parameters. However, if one introduces a fourth free parameter, the analytical treatment of the dynamics is much more involved. The Fokker-Planck equation with just three parameters is an ideal starting point to understand the dynamics governing species-rich ecosystems in a simplified fashion. Indeed, the agreement between empirical data and the macro-ecological properties predicted by the model is consistent with the simplifying assumption that tropical forests are close to their steady state. In addition, because the model is not only neutral, but also non-interacting, one may speculate that surviving species are those that were able to reach a steady state of coexistence by minimizing interspecies  interactions.

Further insights into the nature of stochasticity can be achieved by writing down Eq. (\ref{eq:multi-FP2}) in the equivalent Langevin form, which is an equation for the state variables themselves. Within the It\^{o} prescription \cite{VanKampen1992}, Eq. (\ref{eq:multi-FP2}) is equivalent to the following Langevin equation
\begin{equation}
\dot{x}_i(t)= A_i(\mathbf{x})  + \sum_{j=1}^{S}b_{ij}(\mathbf{x})\xi_j(t)\ ,
\label{Langevin}
\end{equation}
where $ \xi_j(t) $ is Gaussian white noise with a zero mean and correlation $\mean{\xi_i(t) \xi_j(t')}=\delta_{ij}\delta(t-t')  $, and the matrix $ b(\mathbf{x}) $ is defined by the equation $ B(\mathbf{x})=b(\mathbf{x})b^T(\mathbf{x}) $, where the entries of $ B(\mathbf{x}) $ are given in Eq. (\ref{eq:FP-B}). This shows that $ b(\mathbf{x}) $ may be thought of as the ``square root'' of $ B(\mathbf{x}) $. However, because any orthogonal transformation of the noise $ \xi_j(t) $ leaves  the mean and the correlation unchanged, the matrix $b(\mathbf{x})$ is not uniquely determined.

For the present model, the Fokker-Planck Eq. (\ref{fokker}) is equivalent to
\begin{equation}
\dot{x}(t)=b- \frac{x(t)}{\tau}  +\sqrt{D x(t)} \xi(t)
\label{langevin}
\end{equation}
\noindent where we have instead assumed that the time correlation is $\mean{\xi(t) \xi(t')}=2\delta(t-t')$. Eq. (\ref{langevin}) shows how demographic stochasticity accounts for random ecological drift. In fact, given the relatively large number of individuals of any species, one expects that the detailed nature of the stochastic noise is not important, such that fluctuations are proportional to $\sqrt{x(t)}$ due to the central limit theorem \cite{VanKampen1992}. Thus, the multiplicative noise in the Langevin equation is not introduced \textit{ad hoc} and can be justified on the basis of more general considerations.

\subsection{Boundary conditions}
The stochastic dynamics described by Eq. (\ref{langevin}) governs the evolution of the population when it is strictly positive. Assuming that $b>0$ and starting at some positive value, the process cannot reach negative values because the drift brings it to a positive value $b\de{t}$ in the following time interval $\de{t}$. Therefore, the process is always non-negative with the origin acting as a singular boundary, which must be specified on the basis of ecological considerations, i.e. one should define what happens when the random ecological drift occasionally leads to a vanishing population. The most frequent ecological situations are given by the following boundaries

\begin{itemize}

\item We use reflecting boundary conditions at $x=0$ to describe ecological communities in which a vanishing population is immediately replaced by a new individual (belonging to either a new or old species). For instance, this may happen when an ecosystem is coupled to a meta-community that extends across large spatial scales. With these boundary conditions, one can describe ecosystems in which biodiversity is continuously sustained by the net immigration of new individuals of new/old species. This prevents an ecosystem from becoming extinct and can finally achieve a non-trivial steady-state. In the case of the model defined by Eq.(\ref{fokker}), reflecting boundaries are obtained by setting the flux of the probability distribution as zero at $x=0$, although there are some equations for which such boundaries cannot be arbitrarily set, such as Eq.(\ref{fokker}) \cite{Feller}.

\item A second possibility arises when a community can lose individuals without any replacement or a net emigration flux of individuals from the ecosystem exists. One can then describe the system by introducing absorbing boundary conditions at $x=0$. These constraints force ecosystems to march inexorably to extinction and the final steady-state corresponds to the complete extinction of species, i.e. $\lim_{t\rightarrow\infty}P(x,t)=\delta(x)$. Usually, absorbing boundaries are obtained by setting the probability distribution at $x=0$ equal to zero, yet some equations do have norm decreasing solutions, such as Eq. (\ref{fokker}), \cite{Feller} which do not vanish at $x=0$. \end{itemize}

\noindent It is possible to define other kinds of boundaries, according to different ecological behaviors when the population consists of just a few individuals. However, the reflecting and absorbing boundaries capture the most interesting situations in ecology.

As we have alluded to above, the reflecting and absorbing solutions of the Fokker-Planck equation (\ref{fokker}) have rather unusual properties. This is essentially due to the fact that the diffusion term, $Dx$, vanishes at $x=0$. Obviously, it is always possible to rewrite a vanishing diffusion term at zero into a non-vanishing one with a suitable change of variables, yet the same change of variables leads to a singular drift term at the boundary. Thus, equations with a vanishing diffusion term or with a singularity of the drift term at the boundary in general share similar unusual properties. In fact, there exist regions in the parameter space of Eq. (\ref{fokker}) in which one cannot arbitrarily fix a boundary at zero, and the solution is completely defined once the initial condition is prescribed. This is because the probability of accessing the origin depends on the values of the parameters. The problem can be studied by introducing the probability to reach the origin for the first time. It is possible to show \cite{Masoliver2012} that this first-passage probability to the origin is zero when $b/D>1$ and therefore, in this case one cannot obtain any solution with absorbing boundaries and the solution can only be reflecting at $x=0$. However, a (unique) norm decreasing solution exists when $0<b/D<1$, which corresponds to the regular absorbing solution but that does not vanish at $x=0$. Also, in the same region of parameters, a reflecting solution  exists with an integrable singularity at $x=0$. All ecosystems that we have studied so far are well described by $b$ and $D$ within this parameter region.

\subsection{Stationary and time dependent solutions}\label{subsec:transient}

An ecosystem described by Eq. (\ref{fokker}) can reach a non trivial steady state when setting reflecting boundary conditions at $x=0$. In this case, the normalized stationary solution is the gamma distribution \cite{VanKampen1992}, i.e.
\begin{equation}
P_0 (x) =
\frac{(D\tau)^{-b/D}}{\Gamma(b/D)}x^{\frac{b}{D}-1}e^{-\frac{x}{D\tau}}
\label{stat}
\end{equation}
where $\Gamma(x)$ is the gamma function \cite{Lebedev}. Note the distribution exists only when $b>0$, i.e. when the ecosystem has a net immigration rate of species. $P_0(x)$ is the pdf of the relative species abundance at stationarity and it gives the probability that a given arbitrary species present in the ecosystem has $x$ individuals. Fitting the empirical RSAs allows one to estimate two combinations of the parameters, i.e. $b/D$ and $D\tau$, with $\tau$ being the characteristic temporal scale to approach stationarity. Interestingly, for $0<b/D\ll 1$ one obtains the continuum approximation of the celebrated Fisher logseries \cite{Fisher1943}, the RSA distribution of a meta-community.

One can analytically solve Eq. (\ref{fokker}) at any time with arbitrary initial conditions, so that the evolution of the population can be traced even far from stationary conditions. The time dependent solution with reflecting boundary conditions ($b>0$) and initial population $x_0$ is 
\begin{eqnarray}
 P_r(x,t|x_0,0) & = & \left(\frac{1}{D\tau}\right)^{\frac{b}{D}}
 x^{\frac{b}{D}-1}e^{-\frac{x}{D\tau}}
 \frac{\left[\left(\frac{1}{D\tau}\right)^{2}x_0 x
e^{-t/\tau}\right]^{\frac{1}{2}-\frac{b}{2D}}}{1-e^{-t/\tau}}\notag\\
&  &\hspace{-1.1cm}
\times\exp\left[-\frac{\frac{1}{D\tau}(x+x_0)e^{-t/\tau}}{1-e^{-t/\tau}}\right]
I_{\frac{b}{D}-1}\left[\frac{\frac{2}{D\tau}\sqrt{x_0 x
 e^{-t/\tau}}}{1-e^{-t/\tau}}\right] .\notag\\
  \label{sol2}
\end{eqnarray}
\noindent where $I_{\nu}(z)$ is the modified Bessel function of the first kind \cite{Lebedev}. Because $I_{\nu}(z)\simeq(z/2)^{\nu}/\Gamma(\nu+1)$ when $z\rightarrow0^+$, for large time intervals Eq. (\ref{sol2}) converges to $p_{0}(x)$ in Eq. (\ref{stat}).
By setting $x=Dy^2/4$ ($y>0$), when $b/D=1/2$ the process can be mapped into the simpler Ornstein-Uhlenbeck process \cite{VanKampen1992} with a boundary (reflecting or absorbing) at $x=0$, and when $b/D=1$ it can be mapped into the (reflecting) Rayleigh process \cite{Gardiner1985}.

%\subsection{Some properties of the process}
From the Fokker-Planck Eq. (\ref{fokker}), it is easy to derive
the evolution of the mean population per species. It is simply
\begin{equation}
\mean{x(t)}= b\tau + (x_0-b\tau)e^{-t/\tau}
\end{equation}
\noindent where $\mean{x(0)}=x_0$ is the initial population per species in the community. The mean number of individuals per species converges to $b\tau$ at stationarity, with standard deviation $\tau\sqrt{bD}$. The auto-covariance, $\kappa(t)=\mean{x(t)x(0)}-\mean{x(t)}\mean{x(0)}$, under stationary conditions is given by
\begin{equation}
\kappa(t)=\int_0^{\infty}\big[xx_0-\mean{x(t)}\mean{x(0)}\big]P_{r}(x,t|x_0,0)
P_{0}(x_0)\de{x}\de{x_0}
\end{equation}
\noindent where $P_{r}(x,t|x_0,0)$ is the reflecting solution in Eq. (\ref{sol2}) and $P_{0}(x_0)$ is the stationary solution in Eq. (\ref{stat}). However, $\kappa(t)$ can be more rapidly obtained from Eq. (\ref{langevin}). In fact, since in the  It\^{o} prescription $\xi(t)$ and $ x(t)$ are uncorrelated, one gets $\mean{\sqrt{x(t)}\xi(t)}=\mean{\sqrt{x(t)}}\mean{\xi(t)}=0$ because we have assumed that $ \mean{\xi(t)}=0 $. Therefore the evolution equation for $\mean{x(t)x(0)}$ becomes $\partial_t\mean{x(t)x(0)}=b\mean{x(0)}-\mean{x(t)x(0)}/\tau$, whose solution is $\mean{x(t)x(0)}=(b\tau)^2+bD\tau^2e^{-t/\tau}$. Hence, under stationary conditions, $\kappa(t)=bD\tau^2e^{-t/\tau}$ and therefore, the time correlation function is $e^{-t/\tau}$. This shows that $\tau$ is the characteristic relaxation time of the process, the temporal scale upon which the system at stationarity recovers from a small perturbation.

\subsection{The species turnover distribution} \label{stdref}

According to the NT, the turnover of ecological communities reflects their continuous reassembly through immigration/emigration and local extinction/speciation.  Species' histories overlap by chance due to stochasticity, yet their lifetimes are finite and distributed according to the underlying governing process. Non-trivial stationary communities are reached because old species are continuously replaced by new species, bringing about a turnover of species that can be studied and modeled within our framework.

To measure species turnover one usually considers the population of a given species at different times, then studying the temporal evolution of their ratios. For an ecosystem close to stationarity, one can look at the distribution $\mathcal{P}(\lambda,t)$, i.e. the probability that at time $t$ the ratio $x(t)/x(0)$ is equal to $\lambda$, where $x(t)$ and $x(0)$ are the population of a given species at time $t>0$ and $t=0$, respectively. Thus, the species turnover distribution (STD) $\mathcal{P}(\lambda,t)$ by definition is
\begin{eqnarray}
\mathcal{P}(\lambda,t) & = & \langle \delta(\lambda-x/x_0)\rangle\notag\\
&&\hspace{-1cm} =
\int_0^{\infty}\!\!\!\de{x_0}\int_0^{\infty}\!\!\!\de{x}\;
P(x,t|x_0 ,0)P_0(x_0)\delta(\lambda-x/x_0). \notag\\
\label{delta}
\end{eqnarray}
Here $P(x,t|x_0 ,0)$ is either the reflecting or absorbing solution defined in Eqs. (\ref{sol2}) and (\ref{absol}), and $P_0(x)$ is given in Eq. (\ref{stat}), and $\lambda>0$ and $t>0$. 
%It is worth noting that as a consequence of the definition itself (regardless of any boundary condition), $\mathcal{P}(\lambda,t)$ has an interesting symmetry that can be demonstrated on more general grounds: $\lambda \mathcal{P}(\lambda,t)=\lambda^{-1}\mathcal{P}(\lambda^{-1},t)$ for any $t$, which is a symmetry for $\lambda=1$, i.e. $x=x_0$, a relative maximum for $t\ll \tau$.

\begin{figure}[htbp]
\begin{center}
\includegraphics[width=19pc]{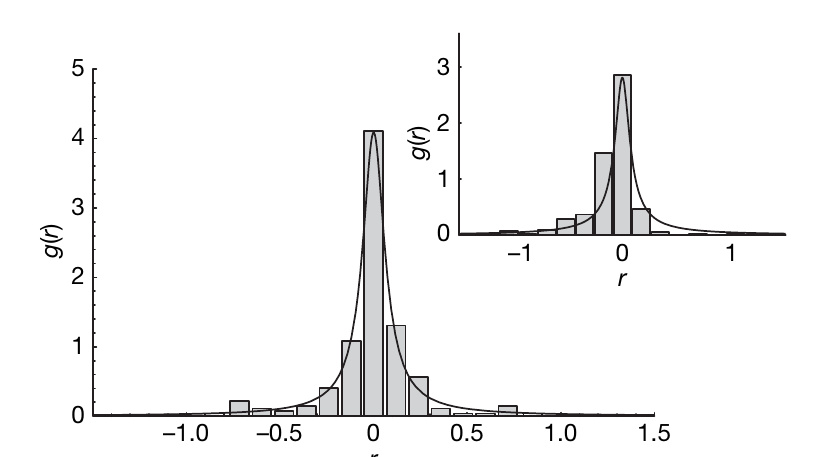}
\caption{[$From$ \cite{Azaele2006}]. STD for the interval 1990-95 in the BCI forest. The main panel shows the results for individuals of more than 10 cm d.b.h., and the inset the results for individuals of more than 1 cm d.b.h. (Center for Tropical Forest Science website). The  black line represents the analytical solution given by Eq. (\ref{delta}).}
\label{STD}
\end{center}
\end{figure}

In Eq. (\ref{delta}), one can use the time-dependent reflecting or absorbing solution according to different ecological dynamics. 
%These possibilities correspond to two distinct types of measures at stationarity.
We should use the reflecting solution when not concerned with the extinction of the species present at the initial time point and especially, when accounting for any new species introduced through immigration/speciation.
% or any old species reappeared after an apparent extinction until $t>0$. 
%The absorbing STD monitors the species present at the initial time point, and does not allow any new species to be introduced by immigration/speciation or the reappearance of a species after extinction. The two distributions are indistinguishable when $t\ll \tau$.
The expression for the reflecting STD can be found in Ref. \cite{Azaele2006}. 
%\begin{widetext}
%\begin{eqnarray}
%\mathcal{P}(\lambda,t) & = & \frac{2^{\frac{b}{D}-1}} {\sqrt{\pi}}
%\frac{\Gamma\left(\frac{b}{D}+\frac{1}{2}\right)}
%{\Gamma\left(\frac{b}{D}\right)}\frac{(\lambda+1)}{\lambda}\frac{\left(e^{t/\tau}
%\right)^{\frac{b}{2D}}}{1-e^{-t/\tau}}\times \notag\\
%& & \hspace{-1cm}\times
%\left(\frac{\sinh\left(\frac{t}{2\tau}\right)}
%{\lambda}\right)^{\frac{b}{D}+1}
%\left(\frac{4\lambda^2}{(\lambda+1)^2e^{t/\tau}-4\lambda}\right)^{\frac{b}{D}+\frac{1}{2}}.\notag\\
%\label{pidilambda}
%\end{eqnarray}
%%\end{widetext}

One can show that it has the following power-law asymptotic behaviour for a fixed $t$
\begin{equation}
\left\{
\begin{array}{lcl}
 \mathcal{P}(\lambda,t) & \approx & \frac{k_1(t)}{\lambda^{\frac{b}{D}+1}}
 \qquad \textrm{for} \quad \lambda \gg 1\\
\vspace{.05 cm}\\
  \mathcal{P}(\lambda,t) & \approx & k_2(t) \lambda^{\frac{b}{D}-1}
  \qquad \textrm{for} \quad \lambda\ll 1
\end{array} \right.
\end{equation}
\noindent where the functions $k_1$ and $k_2$ are independent of $\lambda$. 
%Additionally, note that $\mathcal{P}(\lambda,t)$ depends only on $b/D$ and $\tau$ and therefore, it is not possible to determine all the free parameters of the model by best fitting only the reflecting STD. Also, one needs a further condition that is provided by the relative species abundance in our case, i.e. Eq. (\ref{stat}). 
Customarily, rather than the random variable $\lambda$, ecologists study $r=\log(\lambda)$ that is distributed according to $g(r,t)=e^r \mathcal{P}(e^r,t)$ and that can be compared to empirical data Fig.\ref{STD}.
%Because we suppose that the tropical forest is initially at or near stationarity, we first fit the RSA data with Eq. (\ref{stat}), and then the expression for the STD with Eq. (\ref{pidilambda}) using only $\tau$ as a free parameter even though the data are not statistically significant at the tails.
We obtain an estimate of $\tau$, the characteristic time scale for the BCI forest, which is around $3500 \pm 1000$ yrs (for trees with $> 10$ cm of stem diameter at breast height (dbh)) and $2900 \pm 1100$ yrs (for trees with $> 1$ cm dbh), where the broad uncertainty is due to the fact that the data are sampled over relatively short time intervals. 

These fits not only provide direct information about the time scale of evolution but also, they underline the importance of rare species in the STD. $b/D$ is closely tied to the distribution of rare species and in fact, $P_{0}(x) \sim x^{b/D-1}$ for $x\ll D\tau$. Alternatively, the dependence of the STD on $b/D$ and $\tau$ suggests that at any fixed time $t>0$, rare species are responsible for the shape of the STD.

\subsection{Persistence or lifetime distributions} \label{lifetime-sec}

A theoretical framework to study and analyze persistence or extinctions of species in ecosystems allows one to understand the link between environmental changes (like habitat destruction or climate change \cite{diamond1989,brown1995,Thomas2004,svenning2008,May2010}) and the increasing number of threatened species.

\begin{figure}[htbp]
\begin{center}
\includegraphics[width=19pc]{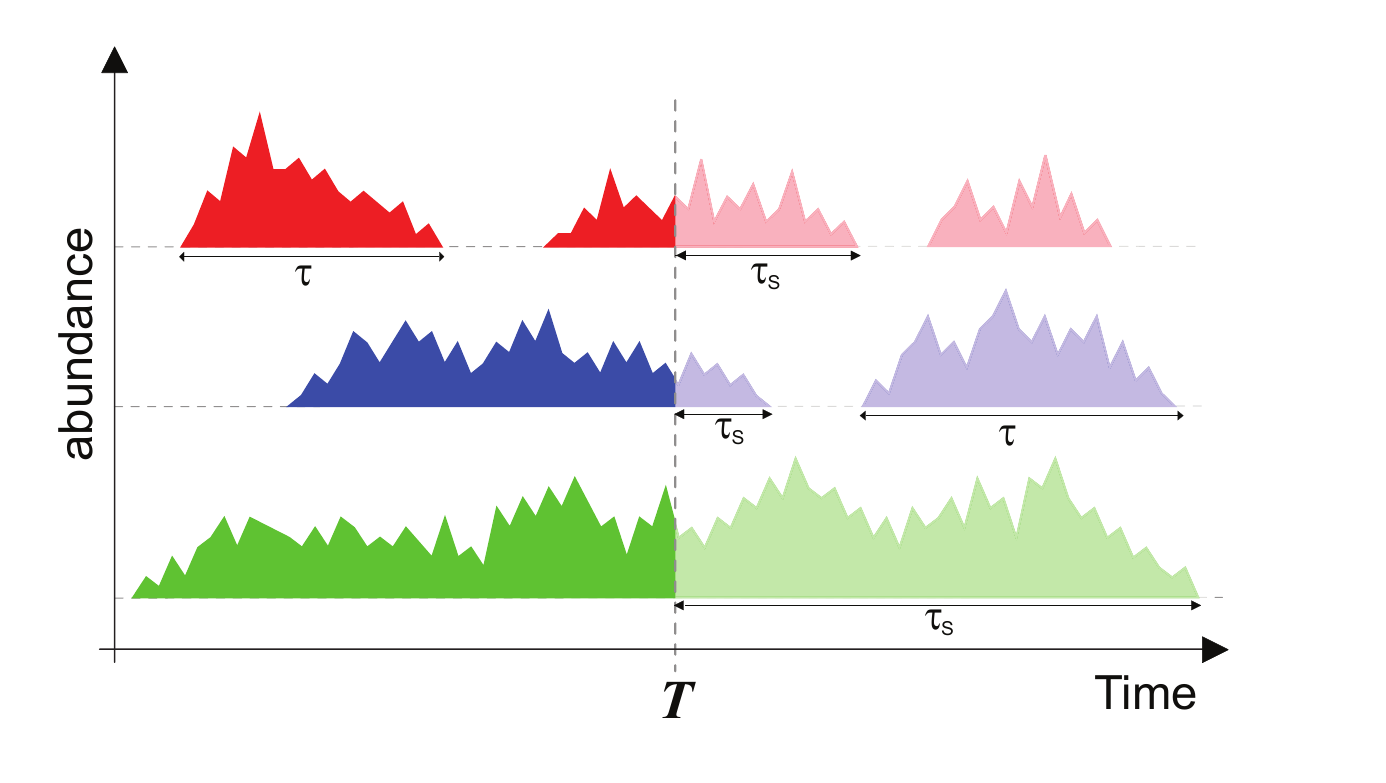}
\caption{[$From$ \cite{Suweis2012}]. Schematic representation of persistence time (or lifetime) of a species $\tau$ and survival times $\tau_s$, defined as the time to local extinction of a species randomly sampled among the observed assemblages at a certain time T.}
\label{SPT-emp0}
\end{center}
\end{figure}

The persistence or lifetime $\tau$ of a species is defined as the time interval between its emergence and its local extinction (see Fig. \ref{SPT-emp0}) within a given geographic region (see~\cite{keitt1998,pigolotti2005species}). In statistical physics, this is known as the distribution of first passage times to zero of the stochastic processes describing the species abundance dynamics. According to the neutral theory of biodiversity, species can span very different lifetime intervals and thus, at a local scale, persistence times are largely controlled by ecological processes like random drift, dispersal and immigration.

\subsubsection{Discrete Population Dynamics}

The simplest baseline model to study the persistence time distribution is a random walk in the species abundance $\vec{n}$, i.e. ME (\ref{1ME}) with $b_n=d_n=c$ and absorbing boundary condition in $n=0$. According to this scheme, local extinction is equivalent to a random walker's first passage to zero and thus, the resulting persistence-time distribution has a power-law decay with exponent $3/2$~\cite{chandrasekar1943,pigolotti2005species}.

A further step in modeling life-time distributions can be made by taking into account birth, death and speciation \cite{Hubbell2001,Volkov2003,Alonso2006,muneepeerakul2008} through a mean field scheme of the voter model with speciation~\cite{Durrett1996,Chave2002} (see description in section \ref{sec:spatial}), i.e. ME (\ref{eq:me-pop}) with birth and death rates given by eqs. (\ref{bM}) and (\ref{dM}) in the large $ J_M $ limit, $b(n)=(1-\nu)n/J_M$ and $d(n)=n/J_M$ for $ n\geq 0 $. The corresponding persistence or life-time distribution is given by $p_{\tau}(t)=-\frac{dP(0,t)}{dt}$, where $P(0,t)$ is the probability that a species has a zero population at time $t$. The asymptotic behavior of the resulting persistence-time distribution (i.e. $p_{\tau}(t)$) exhibits a power-law scaling modified by an exponential cut-off~\cite{pigolotti2005species}:
\begin{equation}
p_{\tau}(t) \ \propto \ t^{-2}e^{-\nu t},\label{eq:SPT}
\end{equation}
for $ t $ greater than some lower cut-off value. Here, $ t $ is measured in units of $J_M$ and $\nu$ is now a speciation rate rather than a probability (Eq. (\ref{eq:SPT}) is derived in appendix \ref{lifetime}). We note that, in this context, $p_{\tau}(t)$ has a characteristic timescale $1/\nu$ for local extinctions determined by the speciation/migration rate. The general case when none of the coefficients is zero and are given by Eqs. (\ref{para}) can also be solved. In this case, $p_{\tau}$ displays a crossover from the $t^{-3/2}$ to the $t^{-2}$ behavior at a certain characteristic time and finally, an exponential decay beyond yet another characteristic time scale \cite{pigolotti2005species}.
It has been shown numerically \cite{Bertuzzo2011,suweis2012species} that for the spatial voter model in dimensions $d=1,2,3$, the exponent of the power-law in Eq. (\ref{eq:SPT}) changes to $t^{-\alpha}$, with $\alpha<2$ depending on the topological structure underlying the voter model. In particular, $\alpha=3/2$ in $d=1$ and $\alpha= 2$ for any $d \geq 3,$ whereas in $d=2$ one gets $p_\tau(t) \propto t^{-2}e^{-\nu t}\ln t $  as shown in \cite{pinto2011} and references therein.

\begin{figure}[htbp]
\begin{center}
\includegraphics[width=19pc]{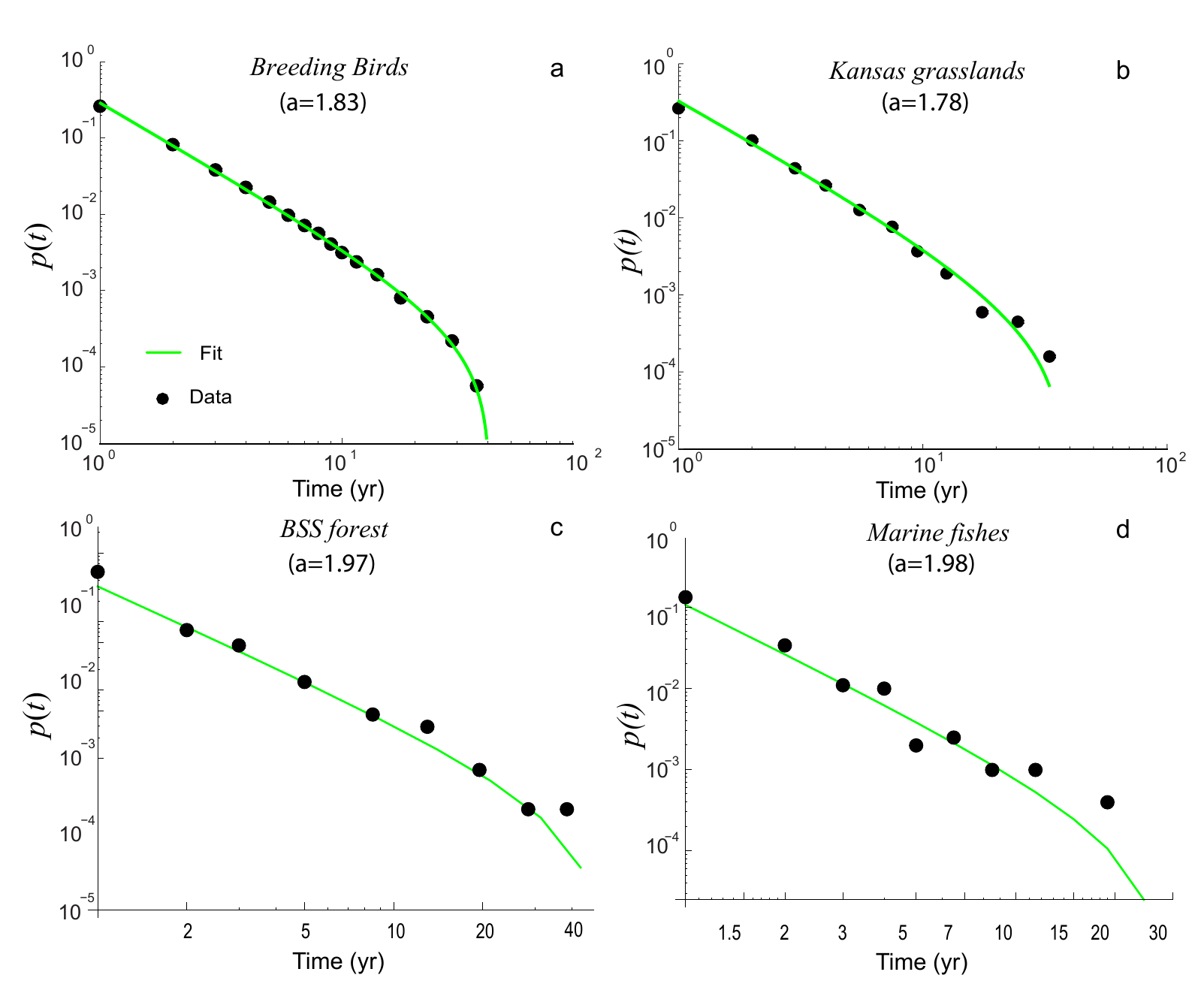}
\caption{[Modified $from$ \cite{suweis2012species}].  Comparison between persistence empirical distributions for (a) North American Breeding birds, (b) Kansas grasslands, (c) New Jersey BSS forest, (d) an estuarine fish community and the corresponding theoretical species persistence times pdfs. The circles and solid lines show  the observational distributions and fit, respectively. The finiteness of the time window $\Delta T_w$ imposes a cut-off in the maximum observable persistence time and thus, only lifetimes where $\tau<\Delta T_w$ have been considered and the theoretical predictions have been adjusted appropriately (appendix \ref{lifetime}). }
\label{SPT-emp}
\end{center}
\end{figure}

We have seen in section \ref{sec:statics} that the RSA pattern does not depend on the biological details of the ecosystem under analysis. Thus, one may wonder if the persistence time distribution is also a "universal" macro-ecological pattern. Indeed, it has been shown \cite{Bertuzzo2011,suweis2012species} that the power-law with an exponential cut-off shape predicted for the persistence time distribution by the NT (see appendix \ref{lifetime}) is common to very different types of ecosystems (see Figure \ref{SPT-emp})

Another interesting and related quantity is the $survival$ $distribution$ $P_{\tau_s}$ defined as the probability that a species randomly sampled from the community at stationarity is still present in the community after a time t. This quantity depends on the initial conditions as $P_{\tau_s}(t)=\int p_{\tau}(t|n_0)P_0(n_0)dn_0$, where $p_{\tau}(t | n_0)$ is the lifetime distribution for a species that initially has a population $n_0$. Assuming that the stationary distribution of population abundances is the Fisher log series, then $P_{\tau_s}(t) \sim t^{-1}$ when $t \ll t^*$, whereas $P_{\tau_s}(t) \sim e^{-t/t^*}$ when $t \gg t^*$ \cite{pigolotti2005species,suweis2012species}. It can be shown that this asymptotic behavior of the survivor distribution is valid regardless of the functional shape of the birth and death rate $b(n)$ and $d(n)$ involved in the ME driving the evolution of $P(n,t)$ \cite{Suweis2012}.
\subsubsection{Continuum limit}

In the continuum limit, a crucial distribution for the analysis of species' extinction is the time dependent solution of Eq. (\ref{fokker}) with absorbing boundaries at $x=0$. We refer to \cite{Feller} for its complete derivation. This probability distribution only exists when $b<D$ and is given by
\begin{eqnarray}
P_{a}(x,t|x_0,0) & = &
\frac{(D\tau)^{-1}}{1-e^{-t/\tau}}\exp\left[-\frac{\frac{1}{D\tau}(x+x_0e^{-
t/\tau})}{1-e^{- t/\tau}}\right]\times \nonumber\\
&  &  \hspace{-1cm}\times\left(\frac{x}{x_0}e^{
t/\tau}\right)^{\frac{b}{2D}-\frac{1}{2}}
I_{1-\frac{b}{D}}\left[\frac{\frac{2}{D\tau}\sqrt{x_0
x e^{t/\tau}}}{e^{ t/\tau}-1}\right].\nonumber\\
  \label{absol}
\end{eqnarray}
Note that (\ref{absol}) is finite at $x=0$ but $\lim_{x\rightarrow0^{+}}P_{a}(x,t|x_0,0)\neq0$.

The lifetime distribution can be calculated analytically using more sophisticated methods through Eq. (\ref{absol}), i.e. $p_{\tau}(t)=-\frac{d}{dt}\int_0^{\infty}P_a(x,t|1,0)dx$. It can be shown that the lifetime distribution calculated in this way displays the same asymptotic behavior as (\ref{eq:SPT}), although the functional form is different \cite{Azaele2006}. In the continuum limit, one can also obtain the mean extinction time\cite{Azaele2006}:
\begin{equation}\label{avertime}
\mean{t}=\tau \int_0^{1}\frac{x^{-b/D}-1}{1-x}\de{x} =
-\tau(\gamma + \psi(1-\beta))
\end{equation}
\noindent which depends only on $0<b/D<1$ and $\tau$. Here, $\gamma=0.577\ldots$ is the Euler constant and $\psi(z)=\Gamma'(z)/\Gamma(z)$ is the logarithmic derivative of the Gamma function \cite{Lebedev}.
Eq. (\ref{stat}) can be used to fit the RSA of various tropical forests yielding $\mean{t}/\tau$ = 1.94, 1.67, 0.67, 0.95 and 1.38 for Yasuni, Lambir, Sinharaja, Korup and Pasoh, respectively (see \cite{Azaele2006}). These time scales are in accord with the estimates of extinction times presented elsewhere \cite{pimm1995} and it is quite interesting that $\mean{t}/\tau$ depends on $b/D$ only, which can be calculated from the steady-state RSA without the need for dynamic data. The values of $ b/D $ obtained from various tropical forests \cite{Azaele2006} suggest that $ \mean{t}\simeq \tau  $, athough this is not built into the model. In general, if $ \tau \gg \mean{t} $, extinction would be much faster than recovery and the ecosystem will not reach a steady state. However, if $ \tau \ll \mean{t} $ the ecosystem would recover from external disturbances very rapidly with respect to the extinction time and therefore, it would be very robust. This would leave little room for the action of evolution. Therefore, the fact that $ \mean{t}\simeq \tau $ suggests that ecosystems at stationarity might be marginally stable --- not so stable that they are frozen in time and not so fragile that they are prone to extinction. From estimates of the model parameters $b/D$, $\tau$ and thus predictions of $\mean{t}$, many biological and ecological features of the ecosystem may be understood.

\section{Neutral spatial models and environmental fragmentation}
\label{sec:spatial}

Space is an essential element to understand the organization of an ecosystem and most empirical observations are spatial. Understanding the features and dynamics of an ecosystem cannot, therefore, be disentangled from the spatial aspects. Spatial models are basically analytically intractable and the main difficulty is due to the fact that spatial models are not equilibrium models~\cite{Grilli2012}.

Spatial effects can be incorporated into the theoretical framework, with the ME of (\ref{eq:me-pop}) remaining formally the same by considering the index $i$ in $n_i$ as a composite index $i=(\alpha,\mathbf{r})$, where $\alpha$ identifies the species and $\mathbf{r}$ identifies its spatial location. The set of all spatial locations $\mathbf{r}$ will be denoted by $\Lambda$. Now the transition rates should take into account dispersal from nearby locations. To simplify, we will use the notation $n_i(\mathbf{r})$, where $i$ indicates one of the $S$ species and $\mathbf{r}$ a spatial location.

At present, no coherent spatial neutral theory exists but rather, there is a collection of models and techniques that can explain some spatial patterns. In this section, we review some of those approaches.

The relationship between the number of species and the area sampled is probably one of the oldest quantities studied in ecology~\cite{Watson1835}. Schoener~\cite{Schoener1976} referred to it as ``one of community ecology's few genuine laws''. The Species-Area relationship (SAR) is defined as the average number of species $\langle S(A) \rangle$ sampled in an area $A$.

\begin{figure}[htbp]
\begin{center}
\includegraphics[width=0.8\columnwidth]{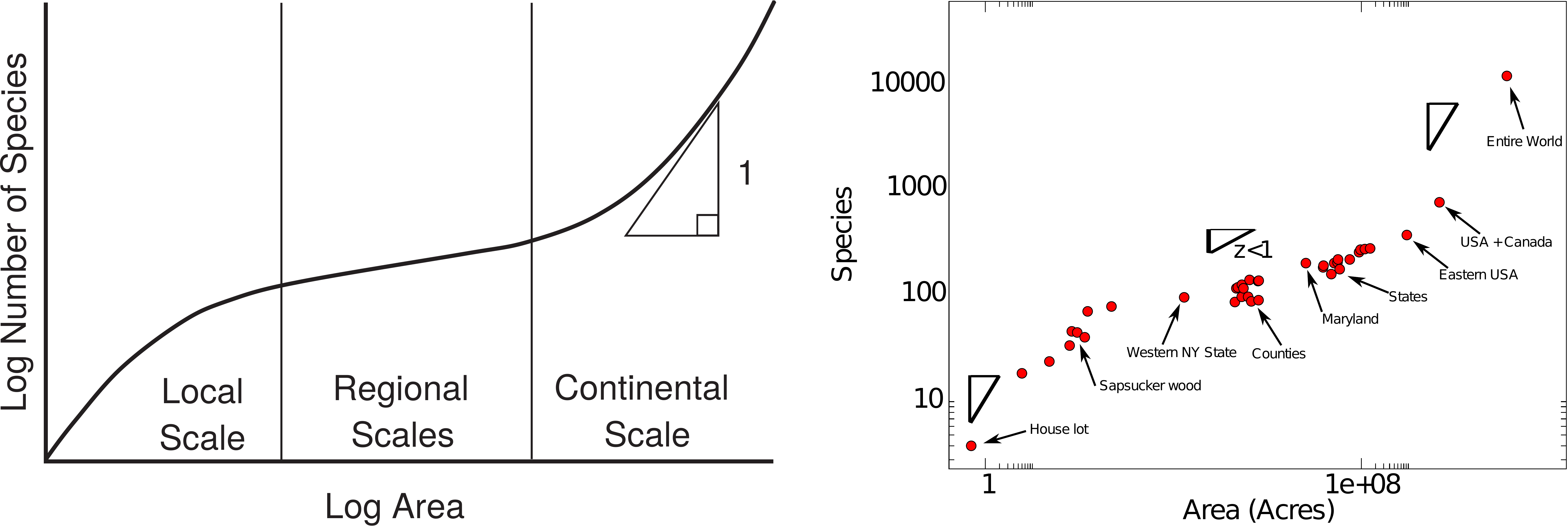}
\caption{Tri-phasic shape of the Species-Area relationship. The left panel shows the three behaviors on different scales. At a local scale the relationship is linear, becoming a power-law relationship at the regional scale and returning to linear at very large intercontinetal scales. The right panel shows the same trend from empirical data~\cite{Preston1960}.}
\label{fig:SAR}
\end{center}
\end{figure}

Arrhenius~\cite{Arrhenius1921} postulated a power-law relationship $\langle S(A) \rangle = c A^z$. Empirical curves shows an inverted $S$-shape~\cite{Preston1960} (see Fig.~\ref{fig:SAR}), with a linear behavior at small and large areas, and a power-law with an exponent $z$ at intermediate scales. This behavior seems to be pervasive and has been reported for distinct ecosystems.

Despite some notable exceptions~\cite{Gould1979}, the value of the exponent $z$ has attracted most attention in studies of SAR. This value of $z$ is far from universal, ranging from $0.1$ to $0.5$ \cite{martin2006} and showing dependence on latitude~\cite{Schoener1976}, body-mass, taxa and general environmental conditions~\cite{Power1972,Martin1981}. The exponent $z$ is interpreted as a measure of biodiversity.
%% https://mannahatta2409.org/info/parameter/131/

Several models tried to reproduce the empirical behavior and a simple but useful assumption involves considering different species as independent realizations of the same process \cite{Coleman1981}. It is important to note that this assumption is stronger than neutrality, because neutrality does not imply independence. Under these assumptions the SAR is given by Eq. (\ref{SAR_Coleman}).

The Endemic-Area relationship (EAR) is defined as the number of species that are completely contained (i.e. endemic) in a given area (see BOX 1). It quantifies the number of species that become extinct when a portion of landscape is destroyed. It is not generally simply related to the SAR~\cite{He2011}. If the species are considered as independent realizations of a unique process, then
\begin{equation}\label{eq:EAR_indep}
\displaystyle
\langle E(A) \rangle = S_{tot}  P_0(A^c)   \ .
\end{equation}
where $S_{tot}$ is the total number of species in the system, $P_0(A^c)$ is the probability that a species is not present in $A^c$, the complement of $A$, and that it is therefore completely contained in $A$.

The $\beta$-diversity is a spatially explicit measure of biodiversity. A simple and useful measure of $\beta$-diversity is the similarity index, i.e. the fraction of common species shared between different locations. It can also be defined as the probability $F(r)$ that two individuals at a distance $r$ are conspecific~\cite{Chave2002}. This quantity may be related to the two point correlation function $G_{ij}(r)$. Under the assumption of translational invariance, this latter is defined as
\begin{equation}
\displaystyle
G_{ij}(r) := \big< n_i(x) n_j(y+r) \big> = \frac{1}{S} \Bigl[ \frac{1}{V} \sum_{x} \sum_{y} n_i(x) n_{j}(y) \delta_{ij} \delta(||x-y||-r) \Bigr]
 \ ,
\label{eq:corr-def}
\end{equation}
where $n_i(x)$ is the number of individuals of the species $i$ in the location $x$, $||x-y||$ is the distance between $x$ and $y$, $V$ is the number of site locations and $\delta(||x-y||-r)$ selects only pairs at a distance $r$. $F(r)$ is the probability that two individuals at a distance $r$ are conspecific, i.e. it is the ratio between the number of pairs of individuals belonging to the same species at a distance $r$ and the total number of pairs of individuals at a distance $r$. Therefore we obtain
\begin{equation}
\displaystyle
F(r) = \frac{  \sum_{i} \sum_{x} \sum_{y} n_i(x) n_{i}(y) \delta(||x-y||-r)  }{
\sum_{i} \sum_{j} \sum_{x} \sum_{y} n_i(x) n_{j}(y) \delta(||x-y||-r)  } \ ,
\label{eq:beta-def1}
\end{equation}
that can be rewritten as
\begin{equation}
\displaystyle
F(r) = \frac{ \sum_{i} \big< n_i(x) n_i(x+r) \big>}{ \sum_{i} \sum_{j} \big< n_i(x) n_{j}(x+r) \big>} =
\frac{ \sum_{i} G_{ii}(r) }{ \sum_{i} \sum_{j} G_{ij}(r) } 
\ .
\label{eq:beta-def2}
\end{equation}
If the spatial positions of different species are independent, then $G_{ij}(r) = \big<n(x)\big>^2 :=\rho^2$,
and by defining $G_{ii}(r)=G_2(r)$,
the $\beta$-diversity reads
\begin{equation}
\displaystyle
F(r) = \frac{ G_{2}(r) }{ (S-1) \rho^2 + G_{2}(r)  } 
\ .
\label{eq:beta-def3}
\end{equation}

Most theoretical work consists of attempts to relate these ecological quantities with other spatial and non-spatial observable factors. For instance, a typical problem is to calculate the SAR knowing the $\beta$-diversity and the RSA over a global scale. One of the future challenges will be to relate and predict spatial patterns on different scales (upscaling and downscaling), having only local information on one or more patterns.

\subsection{Phenomenological models}

Phenomenological models do not assume any microscopic dynamics but they are rather based on a given phenomenological distribution of individuals in space.

The simplest assumption is to consider individuals at random positions in space~\cite{Coleman1981,Coleman1982}. 

This null model, usually known as \emph{random placement}, can be used to obtain predictions for the SAR and EAR having the RSA or the SAD as an input. Even though this assumption is not realistic, \emph{random placement} turns out to be very useful to capture the relevant aspects of the relationship between RSA and SAR, and it also gives reasonable predictions that can be benchmarked against empirical data. In addition, this assumption also allows direct connections between SAR and EAR to be formulated~\cite{He2011}.

Consider $S(A_0)$ species in a region of total area $A_0$. Species $i$ has an abundance $n_i$ and the $N=\sum_i n_i$ individuals are uniformly distributed at random in the area $A_0$. If we observe a sub-region of area $A$, the probability of observing a particular individual is simply $A/A_0$, while the probability of not observing it is $1-A/A_0$. The probability of not observing species $i$ will then be $(1-A/A_0)^{n_i}$, given the fact that the positions of individuals are independent. We can then obtain the average number of species observed in an area $A$ as
\begin{equation}\label{SAR_rndplc}
\displaystyle
\langle S(A) \rangle = S(A_0) - \sum_{i=1}^S \biggl( 1 - \frac{A}{A_0}  \biggr)^{n_i} \ ,
\end{equation}
where $S(A_0)= S_{tot}$ is the total number of species in the system.

The simple framework of the random placement model also allows the EAR to be calculated. Using (\ref{eq:EAR_indep}) we obtain
\begin{equation}\label{EAR_rndplc}
\displaystyle
\langle E(A) \rangle = \sum_{i=1}^S \biggl(\frac{A}{A_0}  \biggr)^{n_i} \  .
\end{equation}
Despite the simplicity of the approximation, the EAR evaluated using random placement captures the quantitative behaviour of several observed ecosystems~\cite{He2011}.

%\begin{figure}[htbp]
%\begin{center}
%\includegraphics[width=19pc]{halley2010.pdf}
%\caption{[$From$ \cite{Halley2011}]. Time scale of extinctions and predictions obtained with the neutral model.
%The quantity $t_{50}$ is defined as the time after which the number of species becomes half of the initial values after habitat loss.
%The top panel compare this quantity to the prediction obtained using neutral theory, with a good match between the two.
%The bottom panel show the relationship between this time scale and the area after habitat loss, which is well approximated by
%, a power-law relationship between the two with an exponent $\sim 0.65$}
%\label{SAR}
%\end{center}
%\end{figure}

Under random placement assumptions, one can obtain the EAR from the SAR and vice versa. To calculate the EAR in (\ref{eq:EAR_indep}), we calculated the number of species with zero individuals in the area complementary to that of interest. This number is equal to the difference between the number of species in the whole area and the number of species in the complementary area. The complementary area has a non-trivial shape and under general assumptions, this quantity is not easy to calculate. Under the random placement assumption, the number of species in the complementary area is the SAR of the complementary area. The EAR is therefore,
\begin{equation}\label{EAR_from_sar}
\displaystyle
\langle E(A) \rangle = S(A_0) - S(A_0 - A)
 \ .
\end{equation}
In~\cite{He2011}, a careful analysis of the reliability of this extrapolation was presented to predict the empirical EAR and it was shown that the random placement
approximation describes the empirical data well.

One can obtain a closed form expression for the EAR and SAR by starting with a RSA distribution. In the case of a Fisher log-series (see eq.~(\ref{P_fisher})), the SAR reads
\begin{equation}\label{SAR_rndplc_fisher0}
\displaystyle
\langle S(A) \rangle = S(A_0) - \sum_{n=1}^\infty \theta \frac{r^n}{n}  \biggl( 1 - \frac{A}{A_0}  \biggr)^{n} =
 \theta \log \bigl( 1 + \frac{r}{1-r} \frac{A}{A_0}  \bigr)  \ ,
\end{equation}
which follows from the observation that $S(A_0)$ is equal to $-\theta \log( 1 - r )$ of eq.~(\ref{P_fisher}). The same calculation can be performed for the EAR, obtaining
\begin{equation}\label{SAR_rndplc_fisher}
\displaystyle
\langle E(A) \rangle = 
- \theta \log \bigl( 1 - r \frac{A}{A_0}  \bigr)  \ .
\end{equation}

In real ecosystems, individuals are not distributed uniformly in space but rather, due to dispersal limitation, individuals of the same species tend to be clustered. This is confirmed by empirical $\beta$-diversity plots~\cite{Chave2002}. The phenomenological approach of random placement can be generalized to a non-uniform distribution of individuals in space, taking into account the empirically observed spatial clustering of conspecific individuals~\cite{Grilli2012a}. Individuals are distributed in space via a Poisson Cluster Process (PCP). In a PCP, the centers of the cluster of points are uniformly distributed in space. A random number of points are distributed around each center according to a given spatial kernel. The process depends on two distributions: the number of points in each cluster and the spatial kernel. It is possible to show~\cite{Grilli2012a} that these two distributions may be related to the RSA and the $\beta$-diversity. An analytical formula for the SAR and the EAR can therefore be obtained given the RSA and the $\beta$-diversity. This approach reproduces the $S$-inverted shape observed in empirical systems, showing that this shape can be explained simply in terms of spatial correlations of conspecific individuals.

\subsection{Spatial Stochastic Processes}

There are two possible ways to include space in a neutral stochastic model, either implicitly or explicitly.

\subsubsection{Spatially implicit model}

Spatially implicit models are based on the observation that one can relate the sample area $A$ to the total number of individuals $J$~\cite{Hubbell2001}, given that $J=\rho A$, where $\rho$ is the density. One can therefore obtain species-area curves in non-spatial models, looking at the scaling of the number of species $S$ with the number of individuals $J$. Spatial implicit models can thus be thought of  as mean-field ``well-mixed'' models, where one can neglect dispersal limitation.

In a meta-community the number of species $\big<\phi(n)\big>$ of a population $n$ is a Fisher log-series (see sec.~\ref{sec:statics})
\begin{equation}\label{SAR_fisher}
\displaystyle
\big<\phi(n)\big> = \theta \frac{r^n}{n} \  ,
\end{equation}
where $r$ is the ratio of the birth to death rate. The average number of individuals in the meta-community is $J_M = \theta r/(1-r)$ and the expected number of species can be easily computed
\begin{equation}\label{SAR_metacom}
\displaystyle
\langle S(J_M) \rangle = \theta \sum_{n=1}^\infty \frac{r^n}{n} = - \theta \log(1 + r)
= \theta \log( 1 + \frac{J_M}{\theta} )  \  .
\end{equation}
This result corresponds exactly to that found using the random placement in (\ref{SAR_rndplc_fisher}). At small sample sizes, the number of species $J_M \ll \theta$ is equal to the number of individuals $\langle S(J_M) \rangle = J_M$. In other words when small areas are sampled the individuals belong to different species and the number of species grows along with the number of individuals. With larger sample sizes,  the number of species $J_M \gg \theta$, grows logarithmically with the number of individuals. This approach allows one to calculate the SAR directly from the RSA distribution and it is clearly applicable to any RSA distribution.

Real ecosystems are of course spatially explicit, but one might wonder how spatial implicit models or, more generally, models that do not consider space explicitly, are predictive and how their parameters are related to spatially-explicit ones. A way to assess this is to measure the efficacy of non-spatial models in predicting the behavior of spatially explicit models~\cite{Etienne2011}. As expected, non-spatial models have a good predictive power when the dispersal lengths are sufficiently large and they are particularly good in predicting non-spatial patterns such as the RSA.

\subsubsection{Spatially explicit model}\label{VMVsec}

Spatially explicit models are typically defined as birth-death-diffusion processes. A model is fully specified given a ME and can be obtained in several ways, i.e. it is possible to write several different MEs that include space in a neutral model. The ME is not tractable analytically and one has to introduce approximations in order to get analytical results. Spatially explicit models are particularly difficult to solve because of the lack of detailed balance (see sec.~\ref{sec:statics}).

\begin{figure}[htbp]
\begin{center}
\includegraphics[width=29pc]{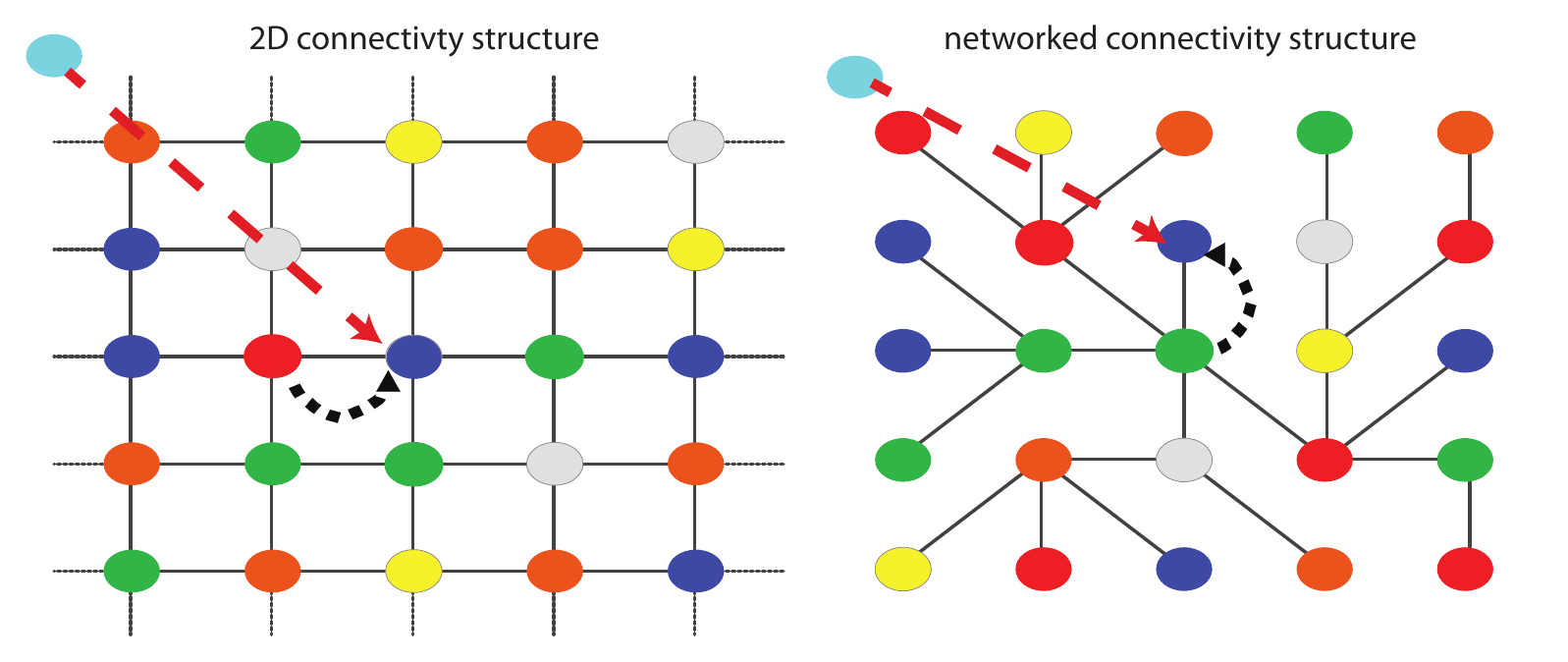}
\caption{Microscopic moves of the Multispecies Voter Model in a $2-$d lattice and with a non-regular network. In the 2D structure, each site is occupied by one and only one individual, whose color represents the species it belongs to. At each time step, one random individual is replaced by a daughter
of one of its neighbors with probability $1-\nu$ (black line). The probability that a speciation event occurs is $\nu$, wherein the individual is replaced by an individual of a new species (red line). In the case of a non-regular network (right panel), the number of neighbors depends on the site considered.}
\label{SpatialVM}
\end{center}
\end{figure}

The voter model~\cite{VoterModel1975} was originally introduced to describe opinion formation, whereby voters are located in a network and each one has one opinion among $q$ possibilities. In ecological applications, voters become individuals and their opinion corresponds to the species they belong to~\cite{Durrett1996}. An individual (voter) is chosen at random and is replaced by a copy of one of its neighbors (see Figure \ref{SpatialVM}). This process has an absorbing state, because when a species disappears there is no way to introduce it again. In the long run, the system is populated by only one species. In order to overcome this problem, one can introduce the possibility of new species entering the system~\cite{Durrett1996,Zillio2005}, a model we will refer to as the multispecies voter model with speciation (MVM).

Consider a lattice of $d$ dimensions (for a typical ecological landscape $D=2$), where $a$ is the lattice spacing with exactly one individual at each site ($a^D$ is the average volume occupied by one individual), and a total number of $N$ individuals. At each time step, one individual chosen at random is removed and is replaced with a copy of one of its neighbors with a probability $1-\nu$, or with an individual of a species not already present in the system with a probability $\nu$. The case $\nu=1$ is trivial, whereas the case $\nu=0$ has an absorbing state, i.e. a state where a single species is present. The MVM is clearly a neutral model, the microscopic dynamics is the same for all species. It is also a zero-sum process, because the total number of individuals $J$ is conserved. The mean field version of the voter model, where the neighbors of each node are all the other nodes, is governed by the ME~(\ref{eq:me-pop}), with the transition rates given by (\ref{bM})-(\ref{dM}).

We want to write an equation for $F(r)$ (the probability that two individuals picked randomly at a distance $r$ belongs to the same species). Assuming translational invariance, one obtains
\begin{equation}
\displaystyle
F(\underline{r})^{t+1} = F(\underline{r})^{t} \Bigl( 1 - \frac{2}{N} \Bigr)
+ \frac{1-\nu}{ d N} \sum_{\mu=1}^{d} \Bigl( F(\underline{r}+\widehat{e}_\mu)^t 
+ F(\underline{r}-\widehat{e}_\mu)^t \Bigr) \ ,
\label{eq:master-voter}
\end{equation}
where $F(\underline{r})^{t}$ is the probability that two individuals separated by $\underline{r}$ at time $t$ belong to the same species. The solution is obtained with the boundary condition $F(\underline{0})=1$.

The stationary solution of (\ref{eq:master-voter}) can be obtained from a Fourier series in the continuum limit, by taking the limit of $a \to 0$, $N \to \infty$ and $\nu \to 0$, and constraining $\gamma^2=2 D \nu / a^2$ to be a constant. Thus, one obtains~\cite{Zillio2005}
\begin{equation}
\displaystyle
F(r) = \frac{c \gamma^{d-2}}{(2\pi)^{d/2}} (\gamma r)^{\frac{2-d}{2}} K_{\frac{2-D}{2}}(\gamma r) \ .
\label{eq:voter-beta-cont}
\end{equation}
where $K_z(r)$ is the modified Bessel function. In one dimension the solution is $F(r) \propto \exp(-r/\xi)$, where the length $\xi=\gamma^{-1}$ is equal to:
\begin{equation}
\displaystyle
\xi = \Biggl( \log \frac{1-\nu}{1-\sqrt{\nu(2-\nu)}} \Biggr)^{-1} \ .
\label{eq:voter-corrlenght}
\end{equation}
The correlation length depends only on the speciation rate $\nu$. It tends to $\infty$ when $\nu \to 0$, i.e. when speciation disappears and individuals of just one species will occupy the entire landscape. The $\beta$-diversity obtained in (\ref{eq:voter-beta-cont}) was shown to be in good agreement with empirical data from different forest censa~\cite{Chave2002}. The prediction could be improved taking into account the Janzen-Connell effect~\cite{Zillio2005}.

The other important quantity that can be deduced with the voter model is the SAR. It is particularly difficult to obtain analytical results, though there are some hints that the behavior can be obtained via a scaling approach~\cite{Zillio2005}, which gives an exponent $z$ independent of the speciation rate $\nu$.

Starting from (\ref{eq:master-voter}) and its stationary solution, one can show that the correlation length scales as $\gamma^{-1}$ or equivalently as
$1/\sqrt{\nu}$. It is therefore natural to choose $A\nu^{d/2}$ as a scaling variable. A second scaling variable can be identified
by noting that $\sum_{\underline{r}} F(\underline{r}) = \big<n^2\big>/\big<n\big> \sim \nu^{-a}$. Therefore, a second scaling variable
can be identified in $n \nu^a$. Using these definitions, one can deduce a scaling form for the SAR
\begin{equation}
\displaystyle
S(A,\nu) = A^z \tilde{S}(A\nu^{d/2})\ ,
\label{eq:sar_scaling}
\end{equation}
where power-law scaling holds when $A\nu^{d/2} \ll 1$ and the exponent $z$ is independent of $\nu$. One can also find a scaling form for the RSA. Denoting $\Phi(A,n,\nu)$ as the fraction of species with $n$ individuals in an area $A$, one obtains
\begin{equation}
\displaystyle
\Phi(A,n,\nu) = n^{-b} \tilde{\Phi}(n \nu^a,A\nu^{d/2})\ .
\label{eq:rsa_scaling}
\end{equation}
By combining the two and fixing the average population per species, one obtains the scaling relationship
\begin{equation}
\displaystyle
a(2-b) = \frac{d}{2}(1-z)
\ ,
\label{eq:hyper_scaling}
\end{equation}
where $b$ is constrained to be lower than $2$. The exponents can be determined numerically in different dimensions~\cite{Zillio2005}.
In particular, a reasonable value of $z=0.3$ in two dimensions is obtained.
%% END SCALING

It is possible to obtain more precise results via extensive numerical simulations. In the case of NT, one can take advantage of the coalescent approach~\cite{Kingman1982,Rosindell2008}. Instead of simulating the stochastic dynamics directly, one can reconstruct the genealogy of the individuals in the sample area by regressing in time. The main value of this method is that there is no need to wait for any transient state to decay and therefore, this approach is much faster than forward dynamics~\cite{Rosindell2008}, as well as allowing infinite landscapes to be simulated.

The coalescent approach has been applied to the MVM with a different dispersal kernel~\cite{Rosindell2007}. Instead of a nearest-neighbor diffusion, once an individual is removed it is replaced by the offspring of another individual with a probability that depends on its distance from the individual removed. In an infinite landscape, the SAR shows the characteristic inverted-$S$ shape. The model depends only on two parameters (up to a choice of the functional form of the dispersal kernel): the speciation rate $\nu$ and the dispersal length $\xi$. The SAR scales as
\begin{equation}
\displaystyle
\big< S(A,\xi,\nu) \big> = \xi^r S(A \xi^{-r},\nu) \ ,
\label{eq:voter-SAR}
\end{equation}
where the exponent $r$ is independent of $\nu$ and very close to $2$ (as expected by dimensional analysis). The exponent $z$ of the power-law regime can be calculated as the derivative of the curve evaluated at the inflection point of the SAR (in log-log scale), i.e. the minimum value of $d \log(S)/ d \log(A)$. Additional simulations~\cite{Pigolotti2009}, obtained with a wider spectrum of speciation rates, suggests a logarithmic relationship
\begin{equation}
\displaystyle
z = \frac{1}{q + m \log(\nu) } \ ,
\label{eq:voter-zexp}
\end{equation}
where $q$ and $m$ are two real parameters,
which confirms the original prediction of~\cite{Durrett1996}. This inverse logarithmic trend seems very robust and it has also been observed in other spatial models~\cite{Cencini2012}. Indeed, it was shown that power-law
dispersal kernels better fit data than other short-ranged kernels \cite{Rosindell2009}, whereby the estimation of the speciation rate corresponding to a given value of the exponent gives much smaller values.

An attempt to connect spatial and temporal patterns can be found \cite{Bertuzzo2011}, where the persistence-time distribution was studied in MVM (see Sec.~\ref{lifetime-sec}). It was shown that the empirical persistence-time distribution is consistent with that predicted by MVM and
\begin{equation}
\displaystyle
p(t|A) \sim t^{-\alpha} e^{- t/\tau(A)} \ ,
\label{eq:voter-persistence0}
\end{equation}
where $\tau$ is the average time of persistence in an area $A$. The exponent $\alpha$ is universal and depends only on the dimension of the system, while the time scale $\tau$ is a function of the area sampled. Empirically, the time scale $\tau$ scales with the area as $A^{\beta}$. In the MVM $\tau = 1/\nu$. Using this fact and that the rate of appearance of a new species is $\lambda = \nu N \sim \nu A$, one can relate these quantities with the SAR. Indeed, the SAR is the product of the rate of appearance of new species and their average persistence time.
\begin{equation}
\displaystyle
S(A) = \lambda \big< t \big> \sim A^{1-\beta(\alpha-1)}    \ .
\label{eq:voter-persistence}
\end{equation}
Here we have assumed that $\big< t \big> \sim \tau^{2-\alpha}$ to obtain a scaling relationship that connects the exponent $z$ and the other exponents: $z = 1-\beta(\alpha-1)$. This allows spatial patterns to be connected with temporal ones in a manner consistent with the empirical patterns shown in Figure~\ref{SPT-emp}.

Other models have been considered in the literature. In a pioneering paper, which received great interest, a ME was proposed that seemed to permit analytical calculations \cite{ODwyer2010}. Indeed, an analytical expression for the SAR was obtained. However, the promised solution was not correct and nor was it an approximation of the actual SAR~\cite{Grilli2012}. The main technical difficulty preventing an analytical solution was that the model was inherently out-of-equilibrium. Detailed balance, which is the condition that makes the calculation of stationary probabilities possible, did not hold~\cite{Grilli2012}. The time is ripe for explorations of this kind. The challenge is to develop new and powerful techniques in non-equilibrium statistical mechanics.

\subsection{Environmental Fragmentation and Habitat Loss}

Resources are not equally distributed and, even over small scales, their distribution in space is far from uniform. This spatial heterogeneity affects the distribution of individuals and species in space, and has clear implications for the conservation of ecosystems. The space within which ecosystems are embedded is often fragmented and a species may be present in only part of the landscape. Moreover, different patches are not independent but they are rather connected via immigration.

In the absence of speciation, the VM~\cite{VoterModel1975} predicts monodominance in dimension $D \leq 2$. Spatial heterogeneity can be modeled as quenched disorder~\cite{Borile2013} and when two species are considered, it is postulated that different locations on a lattice prefer one species over the other. At each site $i$, a binary variable $\sigma_i$ resides that takes values of $\pm 1$, depending on which species is present. Spatial heterogeneity can be  modeled as a quenched external field $\tau_i$, which also takes values $\pm 1$, and the dynamics are fully specified by the transition rates
\begin{equation}
\displaystyle
W[ \sigma_i \to - \sigma_i ] = \frac{1 - \epsilon \tau_i \sigma_i}{2 z} \sum_{j \in \partial i}( 1 - \sigma_i \sigma_j ) \ ,
\label{eq:VM-transrate}
\end{equation}
where $\partial i$ is the set of nearest neighbors of $i$, and $z$ is the size of this set. The quantity $\epsilon$ measures the strength of the preference. The main result from this model \cite{Borile2013} was that this randomness enhances the coexistence of species. Indeed, if $\epsilon$ is larger than $\epsilon_c = \sqrt{2/(2+N)}$, species coexist in any dimension in the limit of a large system.   One can introduce the quantity $\phi = (1/N) \sum_i \sigma_i$ and if coexistence occurs $\big< \phi^2 \big><1$, whereas this quantity tends to one otherwise. It was shown that ~\cite{Borile2013}, for small $\epsilon$
\begin{equation}
P(\phi) \propto \frac{ \exp( - \frac{N}{2} \frac{\epsilon^2}{1-\epsilon^2} \phi^2 )
 }{ (1-\epsilon^2)(1-\phi^2) + 2 \nu }
  \ ,
\label{eq:FluctPhi_VMdis}
\end{equation}
where a small mutation rate ($\nu \ll 2/(2+N)$) was introduced to regularize the solution. If $\epsilon > \epsilon_c$, then the average $\big<\phi^2\big> < 1$, i.e. coexistence is stable.

NT has also been applied to predict the extinction rate of species after habitat loss \cite{Halley2011}. Habitat loss corresponds to a reduction of the area available and therefore, to the total number of individuals. When this area is destroyed, the endemic species suddenly disappear and what follows is a delayed series of extinctions due to habitat loss. The community was modeled as a neutral assembly of species and a typical time scale of extinctions was obtained, along with its dependence on the number of species and the habitat destroyed. The predictions obtained in this way  reproduced the available data of avifaunal extinctions well \cite{Halley2011}.

% \sandro{In this section we nearly do not comment/cite any paper in which we are not co-authors. We should include more such papers for many reasons (e.g. Rosindell \& Cornell, Etienne \& Rosindell, Durrett \& Levin, \ldots), otherwise we'll get harsh reviews.}

\section{Beyond neutrality and open problems}
\label{sec:beyond}

\subsection{Reconciling Neutral and Niche Theory}

The concept of niche is central in classical ecology ~\cite{MacArthur1967,Chesson2000,Chase2003}. An ecological niche is
``the requirements of a species for existence in a given environment
and its impacts on that environment''~\cite{Chase2003} and it describes how an organism or a population respond to changes in resources, competitors and predators. A possible mathematical realization of the concept of niche is the Hutchinsonian niche, which involves a $n$-dimensional hypervolume, where the axes are environmental conditions or resources. A position in this space represents a set of behaviors and traits characterizing a species or a group of individuals. In niche theory much relevance is given to the specific traits of
species and their interdependence. A central concept in niche theory is competitive exclusion, which states that two species cannot occupy the same
niche, as two identical, yet distinct species, cannot co-exist.

The mathematical representation of an ecological community that includes niche aspects typically coincides with the Lotka-Volterra equations.
In this case, the focus is on the properties of the fixed points (or other dynamical attractors) of these systems of equations and the typical problem that is analyzed is their stability, in relation to the parameters and the species present in the system.

The main difference between neutral and niche theory therefore depends on which mechanism plays the main role in shaping ecosystems \cite{ jeraldo2012}. Neutral theory assumes that random processes, such as dispersal, demographic stochasticity, speciation and ecological drift, have a stronger impact on many of the observed patterns than niche differences. Niche theory assumes the opposite, that the quantitatively important processes are related to differences in species and their interdependence.

As we might expect in a real ecosystem, both stochasticity and niche differences play a role, and it is natural to try to quantify how neutral behavior emerges from a niche model. In many cases, niche-based and neutral models yield compatible fits of biodiversity patterns ~\cite{Chave2002b,McGill2003,Mouquet2003,Mcgill2006}, and it's impossible to distinguish between the mechanisms by looking at those patterns. As pointed out ~\cite{Adler2007} neutrality emerges when species have the same or very similar fitness.

Neutrality has often been proposed to emerge under some conditions from models considering niche differences~\cite{Gravel2006,Haegeman2011,Fisher2013}, and neutrality and niche theory were proposed to be the extremes of a more general model~\cite{Gravel2006,Haegeman2011}. In both cases, a Lotka-Volterra equation is introduced to describe community dynamics, while population dynamics is modeled by also taking into account demographic stochasticity and immigration.
By considering different values of parameters, one can move from a scenario where species differences matter a lot and the stable configuration is very close to the solution of the deterministic Lotka-Volterra equation, to a scenario where demographic stochasticity is more important and the community behaves like a neutral community.
A slightly different approach has also been considered ~\cite{Fisher2013}, analyzing a stochastic version of Lotka-Volterra dynamics and quantifying, when stochasticity is varied, the difference between the prediction of the neutral model with the full model. In this case, instead of a continuum of strategies, neutral and niche regimes are two macroscopic phases separated by a phase transition.

A different approach was based on effectively considering niche theory as a
model where per capita death/birth rates are species dependent ~\cite{Borile2012}. In this case, one might expect the difference between species abundances to reflect the differences between these parameters. More precisely, in a neutral scenario, all the species fluctuate around a given abundance value, while when niche characteristics  play a role, each species fluctuates around its own distinct value of abundance.
A third scenario has been proposed~\cite{Borile2012}, wherein the per-capita birth/death rates do not have a monotonic effect on species abundance. The symmetry between species, due to the neutrality of the process, can yet be spontaneously broken. In this case, the stable states are not symmetric in terms of species abundance.

\subsection{Emergent neutrality}

Ecologists have highly criticised NT because of its unrealistic assumptions. The patterns that we have studied so far can in fact be explained without introducing species differences, and this has led some ecologists to oppose NT because it assumes that nature is actually governed by neutral processes, whereas it is not. Clearly, no one believes that nature is truly neutral. The patterns of community ecology are actually generated by a cocktail of processes, and it is both inappropriate and dangerous to consider processes in isolation from a macro-ecological pattern or empirical data set. 

However, it is informative to study whether, how and which ecological processes can drive a community towards a state in which demographic stochasticity and immigration play a crucial role in the face of strong species' differences that are dictated by classical competitive exclusion. Such a state should allow similar species to emerge in the niche space with the ability to co-exist for sufficient time. Indeed, when this problem was studied, it was shown that species can evolve into groups of relatively more similar species that co-exist for very long times \cite{Scheffer2006}. First, a large number of species were placed at random along a hypothetical niche axis, which represents a specific trait, assuming that interspecific competition can be calculated through niche overlap. Running a classical Lotka-Volterra competition model and studying evolution, groups of multiple species were evident that aggregated around similar values in the niche axis and they could co-exist for many generations before the majority of them head towards an inexorable extinction. Eventually, only one species survives from each group, producing the expected pattern of single species equally spaced in the niche space. 

In other words, the niche similarity of species prevents competitive exclusion from swiftly selecting the best competitor among a group of similar species, allowing their co-existence for very long times even though only the superior species will ultimately persist.

This model may be considered as one of the possible steps towards a reconciliation of niche and neutral theories. Species that are initially ecologically non-equivalent, and that therefore behave in a non-neutral fashion, are driven by community and evolutionary processes towards states in which the dynamics may well be better approximated by neutral models over appropriate spatial and temporal scales. More recently, further support for this approach came from showing that the model can produce multimodal RSAs \cite{vergnon2012emergent}. Immigration may also be an important component in the neutral-like behaviour of communities \cite{Gravel2006,Holt2006}. Parasitoids competing for a common species \cite{Bonsall2004} have been used to show that clusters of species separated by gaps emerge along the niche axis,  confirming --- using a quite different approach --- that processes exist that can lead community dynamics to be effectively neutral \cite{purves2005}. 

The basic Lotka-Volterra model has been extended to investigate the possibility that some processes decrease the risk of competitive exclusion so that species lumps are not only transient, but ultimately permanent ~\cite{Scheffer2006}. Density-dependent regulation was introduced that stabilizes the co-existence of species within a group. This approach unfortunately has a drawback that the mechanism introduces a discontinuity in the competition strength of the species \cite{Barabas2013}, which means that unmodeled species differences may be responsible for co-existence in the community. However, it has been shown that more realistic density dependent terms, or even other mechanisms (e.g., migration \cite{vergnon2013repeated}), can eliminate this problem, making the approach more robust \cite{Vergnon2013}.

\subsection{Maximum Entropy Models}
The maximum entropy principle is a useful method to obtain the least biased information from empirical measurements \cite{Jaynes2003}. This powerful tool, borrowed from statistical mechanics, has a wide range of applications \cite{banavar2010applications}, including ecology \cite{banavar2010applications,Harte2011}. 
In its ecological application, the Max Ent principle is an inference method \cite{Chayes1984} used to evaluate the effective strength of interactions among species based on either species abundance data \cite{volkov2009inferring} or simply the presence/absence of the species \cite{azaele2010inferring}. This methodology provides a way to systematically incorporate the most important species interactions into the development of a theory beyond the purely non-interacting case. In addition, the Max Ent principle was implemented as a method to predict biodiversity patterns across different spatial scales using only the information on local interactions \cite{Harte2011,Adorisio2014}. Here we will discuss the extension of Max Ent to the study of spatial biodiversity patterns.
Consider an ecosystem in which $S$ species (belonging to the same trophic level, see BOX 2) live within a given area $A$, divided into $N$ adjacent sites of equal area. Let us assume that there are empirical records of the species contained within each site. From these data one can calculate, for example, the average presence of any species in the ecosystem and the co-occurrence of any pair of species in neighboring sites. When applying Max Ent, we can consider these mean occurrences and co-occurrences as given constraints. We introduce the binary random variable $\sigma_i^{\alpha}$, which records the occurrence of species $\alpha = {1, ... , S}$ at each site $i = {1,...,N}$. If species $\alpha$ is present in plot $i$, then $\sigma_i^{\alpha}=1$, otherwise $\sigma_i^{\alpha}=0$. Therefore the ``state" of any species $ \alpha $ can be characterized by the random vector $ \boldsymbol{\sigma}^{\alpha} $. There is empirical evidence that species belonging to the same trophic level interact weakly  \cite{veech2006probability,volkov2009inferring,azaele2010inferring} and therefore, in a first approximation, one may assume that species occur in a given geographical location independently of one another. Due to such independence, the probability of finding the system in the configuration $ \boldsymbol{\sigma}=(\boldsymbol{\sigma}^1,\boldsymbol{\sigma}^2,..,\boldsymbol{\sigma}^S) $ is thus $P(\boldsymbol{\sigma})=\prod_{\alpha=1}^S p_{\alpha}(\boldsymbol{\sigma}^{\alpha})$, where $p_{\alpha}$  gives the probability distribution of finding a species $ \alpha $ in the configuration $\boldsymbol{\sigma}^{\alpha}$. In order to build the Max Ent model, we will maximize the Shannon's entropy $H=-\sum_{\boldsymbol{\sigma}^1}\cdots\sum_{\boldsymbol{\sigma}^{S}}P(\boldsymbol{\sigma})\ln P(\boldsymbol{\sigma})=-\sum_{\alpha}\sum_{\boldsymbol{\sigma}^{\alpha}}p_{\alpha}(\boldsymbol{\sigma}^{\alpha})\ln p_{\alpha}(\boldsymbol{\sigma}^{\alpha})$, imposing the average occurrence constraint, i.e. $\mean{M}=\sum_\alpha \sum_{\boldsymbol{\sigma}^{\alpha}} p_\alpha(\boldsymbol{\sigma}^{\alpha}) M_\alpha(\boldsymbol{\sigma}^{\alpha})$ with $M_\alpha(\boldsymbol{\sigma}^{\alpha})=\sum_i \sigma_i^\alpha $, and the average co-occurrence constraint, i.e. $
\mean{E}=\sum_\alpha \sum_{\boldsymbol{\sigma}^{\alpha}} p_\alpha(\boldsymbol{\sigma}^{\alpha}) E_\alpha(\boldsymbol{\sigma}^{\alpha})$ with $E_\alpha(\boldsymbol{\sigma}^{\alpha})=\sum_{(i,j)} \sigma_i^\alpha \sigma_j^\alpha$, where $ (i,j) $ indicates two nearest neighbour locations. Both $ \mean{M} $ and $ \mean{E} $ are meant to match the corresponding empirical averages, as calculated from the empirical records of species contained in the region. Maximization \cite{Adorisio2014} provides an explicit expression for the probability, $p_{\alpha},$  of finding a species $ \alpha $ in the configuration $\boldsymbol{\sigma}^{\alpha}$. Thus,

\begin{equation}
 p_{\alpha}(\boldsymbol{\sigma}^{\alpha})=\frac{e^{J_{\alpha}E_\alpha(\boldsymbol{\sigma}^{\alpha})+h_{\alpha}M_\alpha(\boldsymbol{\sigma}^{\alpha})}}{Z_{\alpha}},
\label{maxent}
\end{equation}
where $Z_{\alpha}=Z_{\alpha}(h_{\alpha},J_{\alpha})$ is the partition function. From (\ref{maxent}) one can characterize the spatial biodiversity patterns of the ecosystem,  i.e. calculate the SAR or the EAR \cite{Adorisio2014}. This type of approach may well be suited to infer biodiversity properties of communities over larger spatial scales by upscaling the model results at local scales. However, only a few studies have tackled this problem (see below).

\subsection{Linking different Macro-Ecological Patterns}
Over the last few decades, ecologists have come to appreciate the importance of spatial patterns and processes, and the explicit introduction of space has the potential to revolutionize what we know about natural populations and communities \cite{Tilman1997,storch2007}. It has become apparent that key ecological patterns, such as SAR, RSA and spatial patterns of species distributions and turnover, are intimately intertwined and scale dependent (see Fig.1). However, despite a blossoming of models to address spatial patterns, only a few practical methods have been proposed to link them across different scales. Specifically, spatial approaches \cite{Plotkin2000,Mcgill2003b} lack the needed analytical machinery, whereas most theoretical approaches are not spatially explicit or sufficiently flexible \cite{Harte2008b,volkov2009inferring}. Even neutral theory was conceived to reflect the idealized behavior of natural systems at equilibrium, rather than to reflect non-pristine landscapes produced by environmental change or management \cite{Hubbell2001}. Therefore, a general methodology is required to predict and link these ecological patterns across scales that is sufficiently robust and flexible to allow its application to a range of natural or managed systems. One possible way to tackle this challenging problem is through a theoretical framework inspired by ideas coming from phenomenological renormalization \cite{Azaele2015}. The fundamental assumption at the core of this theoretical setting is that the functional form of the RSA remains the same across all spatial scales, even though the parameters of the curve are likely to vary. Because of this assumption, the spatial dependence of the abundance distribution can be obtained by making the RSA parameters suitable functions of scale. Together with the functional shape of the RSA, the other model input is the spatial pair correlation function (PCF), which describes the correlation in species' abundances between pairs of samples as a function of the distance between them \cite{Zillio2005}. If populations were randomly distributed in space, distinct communities would on average share the same fraction of species regardless of their spatial separation, and therefore, the PCF would not depend on distance. In contrast, in highly aggregated communities correlations in abundance would fall off steeply with increasing distance. The PCF not only measures the rate of turnover in species composition but it also reflects the variation of population clustering across scales, given that the variance in species abundances at any particular scale can be calculated directly from the PCF \cite{Azaele2015}. Therefore, the PCF is related to the spatial species abundance distribution. Thus, the PCF can link the effects of aggregation, similarity decay, species richness and species abundances across scales. Building on the intrinsic relationship among these patterns, the profile of the species area relationships and the species abundance distributions could be predicted across scales when a limited number of fine-scale scattered samples is available \cite{Azaele2015}.

\subsection{Multi-Trophic Ecosystems: Ecological Networks}
Although we are progressing in understanding the suitability and limits of non-interacting and non-spatial models, most neutral models still assume that species interact randomly with each other. However, a network approach to modelling ecological systems provides a powerful representation of the interactions among species \cite{Krishna2008, Bastolla2009, Suweis2013}. Ecological networks may be viewed as a set of different species (nodes) and connections/links (edges) that represent interspecific interactions (e.g., competition, predation, parasitism and mutualism). The architecture of ecological interaction networks has become a bubbling area of research, and it seems to be a critical feature in shaping and regulating community dynamics and structure diversity patterns \cite{Allesina2012}. An important step relevant for multi-trophic systems will be to obtain a general framework within which a network of preferences/disfavors modifies the birth and death rates of different species and can be superposed on neutral models (like the voter model presented in this review).

\section{Conclusions}
\label{sec:concl}
In this paper, we have attempted to describe some of the theoretical frameworks that can be used to understand key issues related to biodiversity and that will serve to address important questions. These frameworks are necessarily elementary and incomplete, yet they have the advantage of being tractable and related to the central issues. Unlike standard approaches to more traditional physics, here the Hamiltonian function or the interactions among the components are completely unknown and even identifying the state variables may sometimes be a non-trivial task. We have a fleeting picture of what an ecosystem is and it is not necessarily in equilibrium. We have little knowledge of the myriad degrees of freedom and their interactions. The real challenge is to discern the most essential degrees of freedom and to develop a framework to understand and predict the emergent ecological behaviour.

In the near future, one needs to investigate how important the species traits are and their interactions as well as environmental effects. At the moment, we do not have a spatially explicit theory that can  predict analytically the most important ecological patterns. This would be important, because it would allow us to understand what drives biodiversity across spatial and temporal scales. Indeed, it is likely that biological processes are not equally important across scales. As evident in particle physics, we might eventually find that ecosystems will need to be described by different effective theories according to the range of scales in which we are interested. Moving towards even more profound questions, a unified theory would ideally explain what determines the total number of species and individuals that live in a given region. Throughout this review, S and N are parameters which have to be provided as input. Currently, we do not have a mechanistic explanation for why one region may be more bio-diverse than another, what sustains biodiversity and how evolutionary pressures sculpt ecological communities.

The basic message of this review is that to resolve these challenging problems, ideas and techniques must be recruited from different disciplines. We are still at the beginning of this adventure. Moving forward is not only important but it is also urgent. The pressures of habitat destruction, pollution and climate change are having highly undesirable consequences on the health of ecological communities. To address practical issues related to conservation biology, we need models that can be used across scales in order to extrapolate information on biodiversity from accessible regions to inaccessible yet important scales. There is plenty of room for ideas that matter, and community ecology can greatly benefit from the contribution of other disciplines, including physics.
%%conclusions here

\section{Acknowledgments}

We have immensely benefited from discussions with Stephen Hubbell, Stefano Allesina, Enrico Bertuzzo, Stephen J. Cornell, Paolo D'Odorico, Andrea Giometto,  
William E. Kunin, Miguel A. Mu\~{n}oz and Andrea Rinaldo. S.S. acknowledges the University of Padova Physics and Astronomy Department Senior Grant 129/2013 Prot. 1634 

\appendix
\addcontentsline{toc}{section}{Appendix A}
\section{Density dependence, and assembling of local communities.}

In this appendix, we present an alternative method to introduce density dependence and how ecosystems emerge by assembling local communities.

First, one can set $b_k(n)=b\cdot(n+\Upsilon_k)$ (for the $k$-th species) and $d(n)=d n$, where $\Upsilon_k$ incorporates both the effects of intra-specific interactions, such as those giving rise to density dependence, and the immigration occurring from a meta-community. This approach can be applied to two distinct ecosystems: coral reefs and tropical forests. 

The steady-state solution of the ME for $P_k(n)$ yields a negative binomial distribution:
\begin{equation}\label{P_fisher_negbin}
P_k(n)=\frac{(1-r_k)^{\Upsilon_k}}{\Gamma(\Upsilon_k)\frac{r^n_k}{n!}\Gamma(n+\Upsilon_k)}
\end{equation}
with a mean $\langle n_k \rangle=r_k\Upsilon_k/(1-r_k)$, where $\Gamma$ is the gamma function.
 
The number of species containing $n$ individuals is given by $\phi_n=\sum_{k=1}^{S}I_{n,k}$, where $I_{n,k}$ is a random variable that is 1 with a probability $P_{k}(n)$ and 0 with a probability $1-P_{k}(n)$. Thus, the RSA is given by

\begin{equation}\label{RSA_densitydependent}
\langle \phi_n \rangle=\sum_{k=1}^{S}I_{n,k}=\sum_{k=1}^S P_{k}(n)=\theta \frac{r^n}{n!}\Gamma(n+\Upsilon),
  \end{equation}
where $\theta=S/[(1-r)^{-\Upsilon}-1]\Gamma[\Upsilon]$ is the biodiversity parameter \cite{Hubbell2001}, and we have dropped the $k$ dependence because of the symmetric hypothesis. For a small $\Upsilon$, the RSA for the communities resembles the Fisher log-series and it does not have an interior mode.

Note that a non-trivial $k$ dependence might arise even under the neutral hypothesis. For example,
one can set $\Upsilon_k = \tilde{m}p_k$, where $\tilde{m}$
is a measure of the immigration rate from the meta-community, in units of the birth rate $b$ and $p_k$ is the fraction of individuals in the surrounding
meta-community belonging to the $k$-th species.
As a result, one can obtain the following RSA for the community of tropical forests.
\begin{eqnarray} \label{rsa}
\langle\phi_n\rangle=\theta\frac{x^n}{n!}\int_0^\infty
\frac{\Gamma(y+n)}{\Gamma(y+1)} e^{-\omega y}
dy\equiv\theta\frac{x^n}{n!}f(n,\omega),
\end{eqnarray}
where $\omega=\frac{\theta}{\tilde m}-\ln(1-x)$.
This approach provides a virtually indistinguishable fit to the empirical data as $\langle\phi_n\rangle=\theta\frac{x^n}{n+c}$ considered earlier with the advantage of having ecologically meaningful parameters.
 
The average number of species observed in the local community is 

  \begin{equation}\label{Saverage}
S_{obs}=\langle S \rangle=S-\langle \phi_0 \rangle=\sum_{k=1}^S(1-r)^{-\Upsilon}=S[1-(1-r)^{\Upsilon}]
  \end{equation}

If the sample considered has $J_L$ individuals, and thus the community dynamics  obeys the zero-sum rule (i.e. it has a fixed total population), then the multivariate probability distribution is 
  \begin{equation}\label{Pmulti}
P(\textbf{n}|J_L)=\mathcal{N}\prod_{k=1}^{S}P_{k}(n)\delta(J_L-n_1-n_2-...-n_S)=\left(
\begin{array}{c}
 J_L+\sum_{k}\Upsilon_k -1 \\
 J_L \\
\end{array}
\right)^{-1}\prod_{k=1}^{S}\left(\begin{array}{c}
 n_k+\Upsilon_k -1 \\
 n_k \\
\end{array}
\right)\delta(J_L-n_1-n_2-...-n_S)
  \end{equation}
that is a compound multinomial Dirichlet distribution where $\mathcal{N}$ is the normalization constant.\\

\textbf{Meta-community Composition.} Now, let us gradually assemble the meta-community of coral reefs by considering it as an assemblage of local communities. Let us start by considering the joint RSA of two local communities, A and B, with $n_A$ and $n_B$ individuals, respectively. The probability that the species has $n$ individuals in the meta-community formed by A and B is\cite{Volkov2007}
  \begin{equation}\label{PmetaAB}
P(n=n_A+n_B)=\sum_{n_A+n_B=n}P(n_A)P(n_B)\propto \frac{r^n}{n!}\Gamma(n+2\Upsilon),
  \end{equation}
where the actual spatial locations of the local reef communities have been neglected (all local communities are well-mixed in the meta-community, i.e. a mean field approximation). The elegant result of the $\Upsilon$'s adding to each other follows ecologically from their interpretation as immigration rates.
For the meta-community, we can introduce speciation with a rate $\nu \ll1$ as we have seen it has a crucial role when $n=0$ (under the assumption of neutrality, the species label of the new species is of no consequence), i.e. $\Upsilon_k=\nu$ and $P(n_k)=\nu r^n/n+O(\nu^2)$.

Extending the calculation of the joint RSA distribution to more and more local communities, it can be shown that the RSA of the meta-community is characterized by an effective immigration parameter $L\Upsilon$, where $L$ is the total number of local communities comprising the meta-community, and it becomes log-normal-like if $L \gg 1$, in agreement with the available data \cite{Volkov2007}.

A Fisher log-series is observed in two limiting cases - in the meta-community in which there are no immigration events and in the very small local community that has a high immigration rate from the meta-community characterized by a Fisher log-series RSA.

\addcontentsline{toc}{section}{Appendix B}
\section{Persistence or lifetime distributions}
\label{lifetime}

In this appendix, we present the derivation for the persistence (or lifetime) distribution that leads to the empirical choice made in (\ref{eq:SPT}).
The master equation we want to solve is 
\begin{equation}\label{ME-B}
\frac{\partial P(n,t)}{\partial t}= b(n-1) P(n-1,t) + d(n+1)P(n+1,t)- \left(b(n)+d(n)\right)P(n,t)\ ,
\end{equation}
where the birth rate is $b(n)= (1-\nu)n/J_M$ with $b(-1)=0$, and the death rate is $d(n)=n/J_M$ and $n\geq 0$. We can redefine the time scale $t \rightarrow J_M t$ so that the factor $1/J_M$ disappears from the birth and death rates. As the initial condition we choose $P(n,t=0)= \delta_{n,n_0}$ with the initial population as $n_0$.
Our goal is to calculate the survival probability defined as
\begin{equation}\label{SP-B}
\mathcal{P}(t|\nu)= \sum_{n\geq 1}P(n,t) = 1 - P(n=0,t)\ .
\end{equation}
Thus, we introduce the generating function
\begin{equation}\label{GF-B}
G(z,t)=\sum_{n\geq 0}P(n,t)z^n \ .
\end{equation}
whose radius of convergence is $\geq 1$. Note that $G(z=1,t)=1$ and $G(z=0,t) = P(n=0,t) = 1 - \mathcal{P}(t|\nu)$. Using Eq. (\ref{ME-B}) the time evolution of the generating function is derived immediately
\begin{equation}\label{GFE-B}
\frac{\partial G(z,t)}{\partial t}= [(1-\nu)z-1](z-1)\frac{\partial G(z,t)}{\partial z} \ .
\end{equation}
with the initial condition $G(z,t=0) = z^{n_0}$.
The previous equation is a linear partial differential equation and it can be solved by standard methods. One introduces a function $Z(\tau)$ satisfying the time evolution equation
\begin{equation}\label{Z-B}
\frac{dZ(\tau)}{d\tau}= -[(1-\nu)Z(\tau)-1](Z(\tau)-1)\ ,
\end{equation}
with a ''final'' condition $Z(\tau=t)=z$. Using Eq. (\ref{GFE-B}), one is led to 
\begin{equation}\label{GZE-B}
\frac{dG(Z(\tau),t)}{d\tau}= 0\ , \forall \tau\ ,
\end{equation}
implying that 
\begin{equation}\label{SolG-B}
G(z,t) = G(Z(t),t) = G(Z(0),0)= Z(0)^{n_0} \ .
\end{equation}
The solution of eq. (\ref{Z-B}), with the chosen ''final'' condition gives
\begin{equation}\label{Z0-B}
Z(\tau=0) = \frac{1-A(z,t)}{1-(1-\nu)A(z,t)} \quad \text{with} \quad A(z,t)=\frac{1-z}{1-(1-\nu)z}e^{-\nu t}
\end{equation}
and thus,
\begin{equation}\label{SolG_bis-B}
G(z,t) = \left( \frac{1-A(z,t)}{1-(1-\nu)A(z,t)}\right)^{n_0}\ .
\end{equation}
Finally, we get the survival probability 
\begin{equation}\label{Surv-B}
\mathcal{P}(t|\nu) = 1 - G(z=0,t) = 1- \left(1 + \nu\left(e^{\nu t} -1\right)^{-1} \right)^{-n_0} \ .
\end{equation}
In the scaling, if we consider the limit of fixed $\nu t$ as $t$ becomes large, we get the scaling form  
\begin{equation}\label{Scaling-B}
\mathcal{P}(t|\nu) = \frac{1}{t} F(\nu t)\ ,
\end{equation}
with 
\begin{equation}\label{F-B}
F(x) = n_0\frac{x}{e^x-1} \ .
\end{equation}
The lifetime distribution is simply given by 
\begin{equation}\label{LTD-B}
p_{\tau}(t)=-d\mathcal{P}(t|\nu)/dt = \frac{1}{t^2} f(\nu t)\ ,
\end{equation}
where the last equality holds in the scaling limit as
\begin{equation}\label{F-B}
f(x) = n_0e^x\left(\frac{x}{e^x-1}\right)^2 \ .
\end{equation}
which for $x\rightarrow 0$ tends to a non-zero constant. As $x\rightarrow \infty$, $f(x)$ decays exponentially, leading to the empirical scaling form of the kind given by (\ref{eq:SPT}).

\addcontentsline{toc}{section}{Appendix C}
\section{ The species turnover distribution with absorbing boundary conditions}
\label{stdabs}

In this appendix we present another temporal pattern predicted by the model defined by (\ref{fokker}) or (\ref{langevin}). 
For an ecosystem at or near stationarity, the model can provide the exact expression for the STD when using the time-dependent absorbing solution of (\ref{fokker}): the so-called the species turnover distribution with absorbing boundary conditions. Unlike the reflecting species turnover distribution, this
distribution corresponds to a new kind of measure that only accounts for
the species present at the initial time, and it does not take into
account any new species introduced by immigration/speciation or
any old species that reappear after their apparent extinction until $t>0$.
This amounts to the selection of a particular sample and the study of the temporal behavior of those selected individuals. The species turnover distribution with absorbing boundary conditions,
$\mathcal{P}^{abs}(\lambda,t)$, can be obtained through formula
(\ref{delta}) where the conditional probability must now be
$p_a(x,t|x_0 ,0)$ defined as in (\ref{absol}). The final
expression is

\begin{eqnarray}\label{stdabesse}
\mathcal{P}^{abs}(\lambda,t) & = & \frac{\sin(\pi
b/D)}{\pi(1-b/D)}\frac{e^{- t/\tau}\left(e^{
t/\tau}-1\right)^{b/D}}{(\lambda+1)^2}\times\notag\\
& \times &{}_{2}F_1\left(1,\frac{3}{2},2-b/D;\frac{4\lambda
e^{-t/\tau}} {(\lambda+1)^2}\right),
\end{eqnarray}

\noindent where $b/D<1$. Note that $\mathcal{P}^{abs}(\lambda,t)$ decays
exponentially to zero at a rate $(1-\beta)/\tau$, regardless of
$\lambda$.

It is noteworthy that the STD's with absorbing and reflecting boundaries are indistinguishable whenever $t\ll\tau$ and $\lambda$ are not too small (for $b/D<1$). Since the BCI data are sampled over relatively short times intervals (at most 10 years), the distributions are almost the same within the interval
$1/2<\lambda<3/2$.

%\nocite{*}%%%%%
\bibliographystyle{abbrv}
\bibliography{NT_rmp-1}

\end{document}